\documentclass[12pt]{article}
\usepackage[hscale=0.8,vscale=0.8]{geometry}
\usepackage{amsmath}
\usepackage{amsfonts}
\usepackage{booktabs}
\usepackage[title]{appendix}
\usepackage{algorithm}
\usepackage{changepage}
\usepackage{algpseudocode}
\usepackage{mathtools}
\usepackage{pdflscape}
\usepackage{capt-of}
\usepackage{cite}
\usepackage[hyphens]{url}
\usepackage{graphicx}
\usepackage{caption}
\usepackage{subcaption}
\usepackage{multirow}
\usepackage{flowchart}
\usetikzlibrary{arrows.meta}
\usepackage{hyperref}
\newcommand{\ra}[1]{\renewcommand{\arraystretch}{#1}}
\newcommand{\beq}{\begin{equation}}
\newcommand{\eeq}{\end{equation}}
\newcommand{\pdd}[2]{\frac{\partial{{#1}}}{\partial{#2}}}
\newcommand{\nhgactfuncheight}{3.5cm}
\newcommand{\nhgactfuncwidth}{0.48\linewidth}
\newcommand{\nhghaloesheight}{3.7cm}
\newcommand{\nhghaloeswidth}{0.19\linewidth}
\newcommand{\nhghalfwidth}{0.48\linewidth}
\newcommand{\nhgtotalheight}{4cm}
\newcommand{\nhglossheight}{4cm}
\newcommand{\nhglosswidth}{0.48\linewidth}
\newcommand{\nhgcnnwidth}{0.48\linewidth}
\newcommand{\nhgcnnheight}{4.0cm}
\newcommand{\nhgappwidth}{0.24\linewidth}
\newcommand{\nhgappheight}{2.9cm}
\newcommand{\nhghistowidth}{0.48\linewidth}
\newcommand{\nhghistoheight}{4cm}

\begin{document}
\title{Solving an elastic inverse problem using Convolutional Neural Networks}
\author{Nachiket Gokhale\footnote{The author is very grateful to Paul Barbone, Ph.D. (Professor, Mechanical Engineering, Boston University, Boston, MA, USA.) for patiently answering many questions about finite elements and for his constructive comments on an earlier version of this document. Conversations with Arnab Majumdar, Ph.D. (ArcVision Technologies, Kolkata, India), Michael Richards, Ph.D. (Assistant Professor, Department of Biomedical Engineering, Kate Gleason College of Engineering, Rochester Institute of Technology, Rochester, NY, USA) and Mandar Kulkarni, Ph.D. (Houston, TX, USA) are gratefully acknowledged and appreciated.}, Ph.D.\\nachiket.gokhale@arcvisions.com{\footnote{Alternate email: gokhalen@gmail.com}}\\ArcVision Technologies, Kolkata, India.}
\date{\today}
\maketitle
\abstract{We explore the application of a Convolutional Neural Network (CNN) to image the shear modulus field of an almost incompressible, isotropic, linear elastic medium in plane strain using displacement or strain field data. This problem is important in medicine because the shear modulus of suspicious and potentially cancerous growths in soft tissue is elevated by about an order of magnitude as compared to the background of normal tissue. Imaging the shear modulus field therefore can lead to high-contrast medical images. Our imaging problem is: \textit{Given a displacement or strain field (or its components), predict the corresponding shear modulus field}. Our CNN is trained using 6000 training examples consisting of a displacement or strain field and a corresponding shear modulus field. We observe encouraging results which warrant further research and show the promise of this methodology.}
\section{Introduction}
\begin{figure}[!h]
   \centering
    \includegraphics[totalheight=3cm]{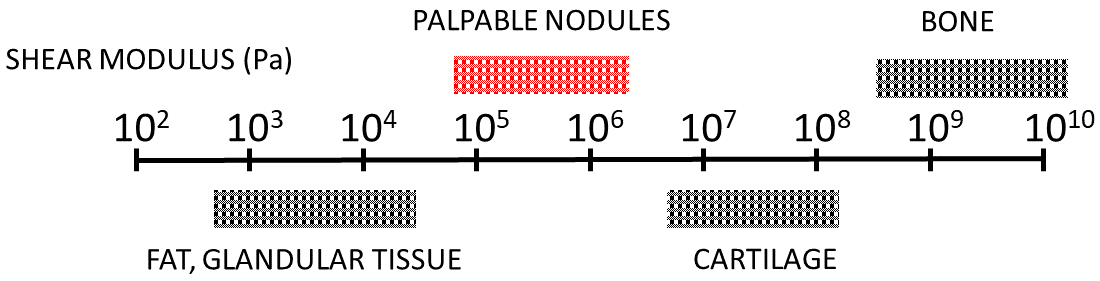}
  \caption{\label{fig:shearmod} Shear moduli of different types of body tissue. Adapted from figure (1) in \cite{paper:sarv1998}.}
\end{figure}

The shear modulus of palpable nodules (which can be thought of as abnormal and potentially cancerous growths in soft tissue) is approximately an order of magnitude higher than the stiffness of the background of normal glandular tissue \cite{paper:sarv1998}. See also figure (\ref{fig:shearmod}). It follows then, that imaging the shear modulus field of soft tissue results in a high-contrast imaging method because the elevated shear modulus of suspicious growths will stand out clearly against the lower shear modulus of the background of normal tissue. Elasticity Imaging (EI) is a broad term that refers to methods which image the shear modulus (or other mechanical properties) of soft tissue in various ways. See \cite{paper:gao1996,paper:parker2010,book:alamgarra2019,bookchap:oberaibarbone2019} for comprehensive reviews of the field.
\begin{figure}[!h]
   \centering
    \includegraphics[totalheight=5cm]{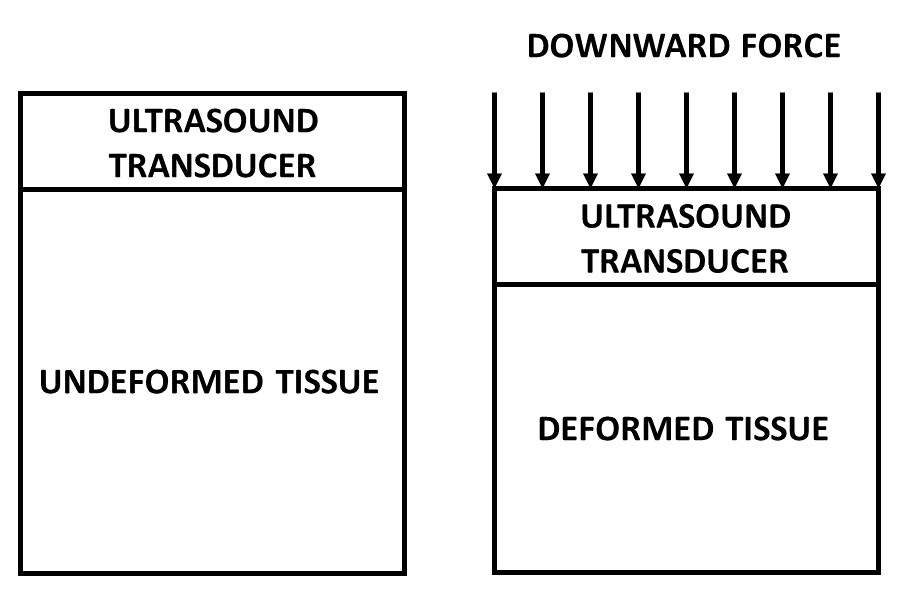}
  \caption{\label{fig:prepostimage} Schematic figure showing medical image acquisition when soft tissue is being deformed using ultrasound imaging. The image taken on the left is referred to as the \textit{pre-deformation} image and the image on the right is the \textit{post-deformation image}.}
\end{figure}
\subsection{Steps involved in elasticity imaging}
Elasticity Imaging typically consists of the three steps of image acquisition, image registration, inverse problem solution. These steps are discussed in the following sections.
\subsubsection{Image acquisition} Images of soft tissue undergoing deformation due to applied excitation are acquired using various medical imaging modalities such as ultrasound or magnetic resonance imaging. While time dependent images can be acquired, we shall consider here only two images: a \textit{pre-deformation image} acquired before force is applied and a \textit{post-deformation image} acquired after force is applied. This process is shown in figure (\ref{fig:prepostimage}) for ultrasound imaging. Also see figure (2) in \cite{paper:konofagou2004}.
\subsubsection{Image registration} 
\begin{figure}[!h]
  \centering
  \includegraphics[totalheight=4cm]{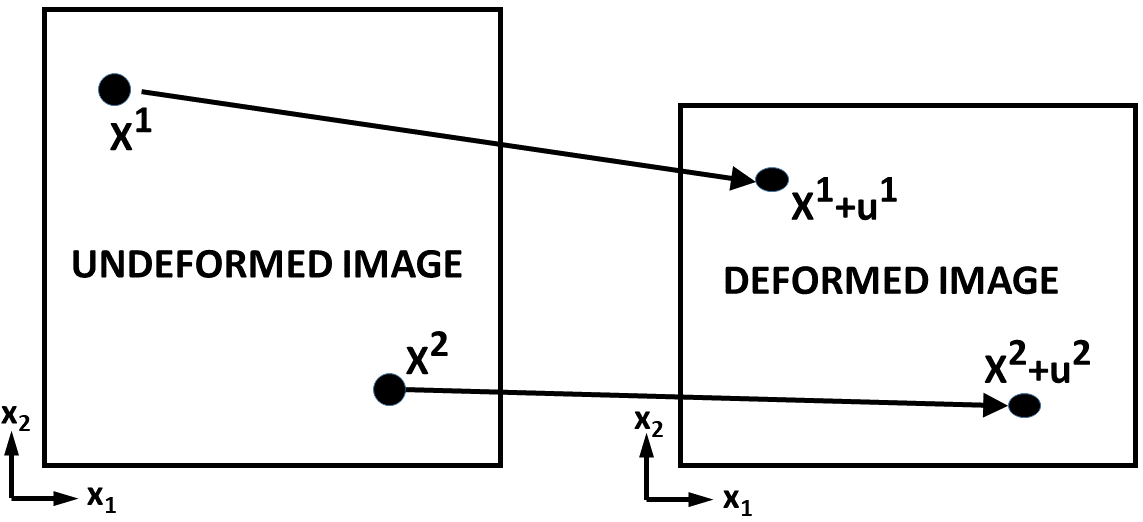}
  \caption{\label{fig:registschematic} A schematic figure of image registration. For points $X^1$ and $X^2$ in the pre-deformation image on the left, we aim to find their location in the post-deformation image on the right by finding the displacements $u^1$ and $u^2$. Doing this for every point in the pre-deformation image yields a displacement field.}
\end{figure}

The goal in this step is to find a map which transforms the pre-deformation image into the post-deformation image. For every point in the pre-deformation image we aim to find its location in the post-deformation image typically by matching image intensity. See figure (\ref{fig:registschematic}). This gives us the \textit{displacement field} between the two images which is often referred to as the \textit{measured displacement field}.

Differentiating the displacement field with respect to spatial coordinates yields the strain field. If $u_x(x,y)$ and $u_{y}(x,y)$ are the $x$ and $y$ components of the displacement field, then the strain field is given by equation (\ref{eqn:straindef}). $\epsilon_{xx}$ and $\epsilon_{yy}$ are referred to as \textit{axial strains}. $\epsilon_{xy}$ is the \textit{shear strain}.
\beq
\label{eqn:straindef}
\epsilon_{xx} = \pdd{u_{x}}{x}, \qquad \epsilon_{yy} = \pdd{u_{y}}{y}, \qquad \epsilon_{xy} = \frac{1}{2}\Big(\pdd{u_{x}}{y} + \pdd{u_{y}}{x}\Big).
\eeq
See \cite{paper:richards2009,paper:gokhale2004,paper:pellot-barakat2004} for minimization based approaches for computing the displacement field. See \cite{paper:ophir1991,paper:ophir1996,paper:alam1998} and references therein for cross-correlation based approaches. In our work, we generate displacement fields from known shear modulus fields using the finite element method (FEM) \cite{book:hugheslinear,book:fishbelytschko} and add an appropriate level of additive white Gaussian noise to mimic the noise when displacement fields are computed from experimentally acquired images. 
\subsubsection{Inverse problem solution} The goal in this step is to infer the spatial distribution of the shear modulus from the displacement field. This is called an \textit{inverse problem} because the classical boundary value problem in linear elasticity (referred to as the \textit{forward problem}) is to determine the displacement field given the shear modulus field, the Poisson's ratio field and suitable boundary conditions. See \cite{book:hugheslinear,book:fishbelytschko,book:segelmathcont} for further details on the boundary value problem of linear elasticity and its solution by FEM.

The approaches for inverse problem solution can be divided into two categories: direct and iterative. These are discussed in the subsequent sections.
\subsubsection{Direct approach} Direct approaches involve solving a partial differential equation (PDE) to obtain the spatial distribution of shear modulus directly: see \cite{paper:raghavan1994,paper:barboneadjwt,paper:albocher}. The coefficients of this PDE depend on the measured displacement field. Such approaches are fast and work well when the measured strain field is completely known and has low noise.
\subsubsection{Iterative approach} Iterative approaches \cite{paper:oberai2003,paper:oberaipmb2004,paper:gokhale2008,paper:kallel1996,paper:doyley,paper:goenezen2011} involve guessing\footnote{We may also use CNNs to generate this initial guess.} a distribution for the shear modulus, solving a linear elasticity forward problem to obtain the predicted displacement field, computing the value of an objective function and its gradient (and perhaps Hessian) with respect to the shear modulus. This objective function is a user specified norm of the difference between the predicted displacement field and the measured displacement field. The guessed shear modulus distribution is updated using a suitable optimization procedure such as a modified Newton Raphson scheme as in \cite{paper:doyley,paper:kallel1996} or the BFGS scheme as in \cite{paper:gokhale2008,paper:oberaipmb2004,paper:oberai2003,paper:goenezen2011,paper:richards2009,diss:gokhale2007}. Such approaches are typically slower than direct methods, since they require the solution of approximately $50$ to $100$ forward problems, but have the ability to handle incomplete data (knowing only one component of the displacement or strain field) and complex nonlinear material models such as hyperelasticity \cite{paper:gokhale2008,paper:goenezen2011}.
\subsubsection{Solving the inverse problem with CNNs}
\begin{figure}[!h]
   \centering
    \includegraphics[totalheight=4cm]{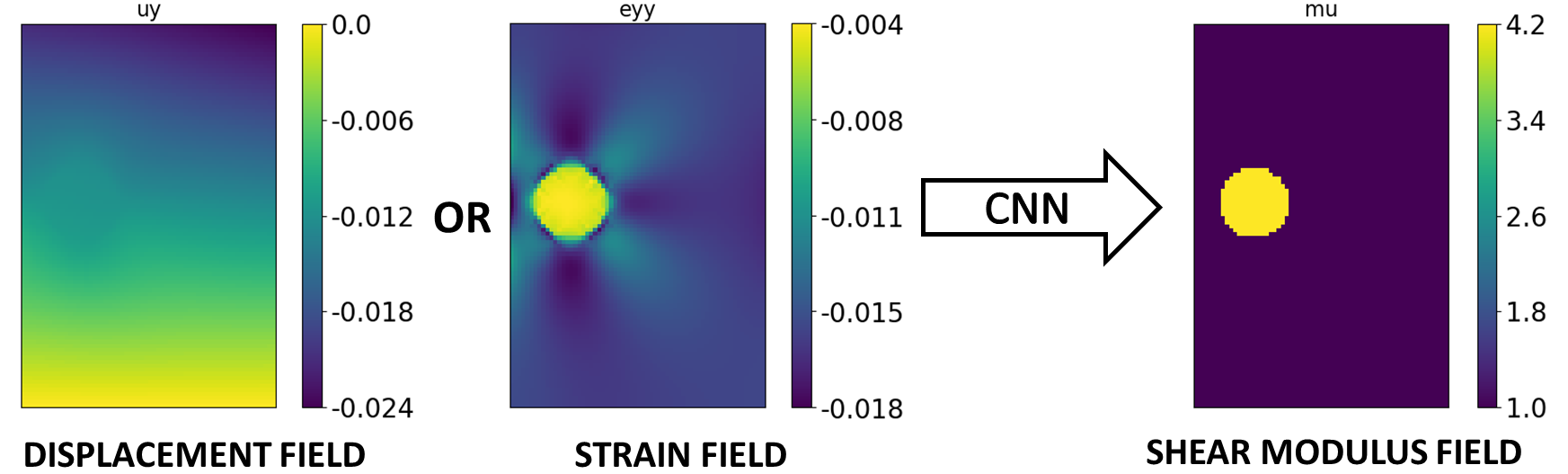}
  \caption{\label{fig:schematic_inv} Solving the inverse problem using CNNs. The CNN maps the displacement or strain field (or components thereof) to a shear modulus field.}
\end{figure}

Conceptually, the solution of the inverse problem using a CNN is shown in figure (\ref{fig:schematic_inv}).
We believe that solving the inverse problem with CNNs can combine some of the best characteristics of the direct and iterative approaches. The CNN based approach can yield a quick answer (once time has been spent up front to train the CNN), can accommodate complex constitutive relations, can work with incomplete data (e.g. only a single component of a displacement field) and can handle noisy data.

On the other hand, there are some disadvantages. We think that if data which is unlike what the CNN has been trained on is seen, then the performance of the CNN will likely degrade. This is unlike traditional inversion schemes. We investigate this in sections  (\ref{sect:resultscnn3}) and (\ref{sect:resultscross}). We also note that CNNs do not predict a perfect result in the absence of noise, unlike traditional direct or iterative methods.

\section{\label{sect:probsetup}Data generation for CNNs}
\begin{figure}[!h] 
   \centering
    \includegraphics[totalheight=5cm]{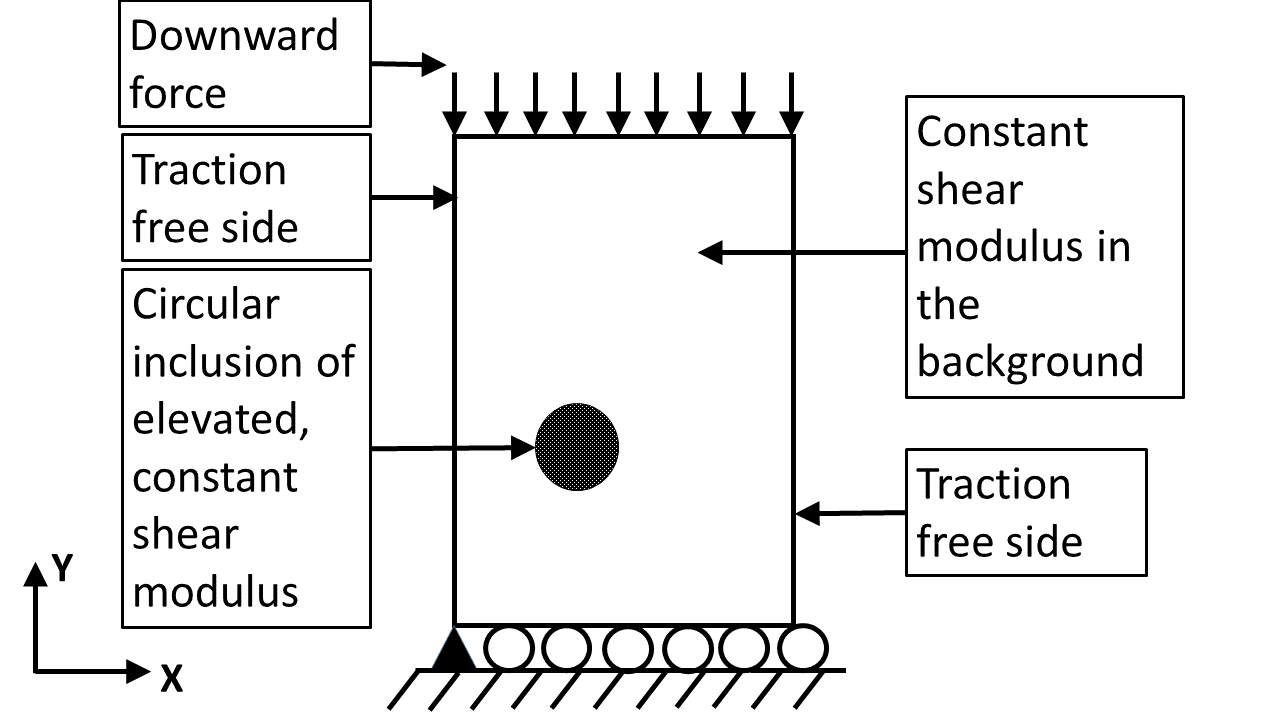}
  \caption{\label{fig:bc}Boundary conditions and material properties used in this work. }
\end{figure}
The displacement or strain field data required for the CNN is generated by solving the boundary value problem of isotropic, almost incompressible, plane strain, linear elasticity \cite{book:segelmathcont} with finite elements \cite{book:hugheslinear,book:fishbelytschko} using a solver named FyPy (\textbf{Fy}nite Elements in \textbf{Py}thon) \cite{misc:fypy}. Both displacements and material properties are interpolated bilinearly. The problem geometry is shown in figure (\ref{fig:bc}). The length (in the $x$ direction) is $1.0$ unit. The breadth (in the $y$ direction) is 1.5 units. There are $65$ nodes in the $x$ direction and $97$ nodes in the $y$ direction. Both degrees of freedom are constrained at the pin and only the $y$ degree of freedom is constrained at the rollers. The background shear modulus is $\mu_{back}=1.0$ unit. Since soft tissues are almost incompressible we set the Poisson's ratio to $0.49$ everywhere which renders the elastic medium almost incompressible. This incompressibility causes a numerical problem with the FEM solver which is called mesh locking \cite{book:hugheslinear}. Mesh locking typically manifests itself as unphysically low displacements for a given traction. We use selective reduced integration \cite{book:hugheslinear} to avoid this problem. The shear modulus of inclusions is a constant and is a random number ranging from $\mu_{min}=2.0$ units to $\mu_{max}=5.0$ units. There are no homogeneous examples. The radius of the inclusion is a random number ranging from $0.05$ units to $0.15$ units.

Solving a single training example takes approximately $5$ seconds. See section (\ref{sect:computresour}) for details of computational resources used. $10000$ displacement and strain images are generated and are split into $6000$ training examples, $2000$ validation examples and $2000$ test examples. The input data is scaled by the absolute maximum over all channels, computed over the training examples only.

When noisy data is used, we add zero mean additive Gaussian noise in the strain or displacement data such that the signal to noise ratio ($\text{SNR}_{\text{dB}}$) is 40dB according to equation (\ref{eqn:snr}). We use 40dB noise because it has been used in prior work \cite{paper:doyley}. We do not train CNNs with noisy data. We always train CNNs with noiseless data. We make predictions by supplying noisy data to the CNNs as an input. 
\begin{subequations}
\begin{align}
  \text{SNR}_{\text{dB}} &= 20\log_{10}{\Bigg (}\frac{\|\text{signal}\|_{2}}{\|\text{noise}\|_{2}}{\Bigg )} \label{eqn:snr} \text{ where, }\\
  \|\mathbf{x}\|_{2} &= \sqrt{x_{1}^{2}+\cdots+x_{n}^2}  \text{ is the Euclidean norm.}   \label{eqn:eucnorm}
\end{align}
\end{subequations}  
\subsection{\label{sect:computresour} Computational resources used}
In this work, we used Python 3.8 to write the finite element solver \cite{misc:fypy} and associated scripts \cite{misc:mlscripts} for generating and post-processing the data. We used the "spsolve" sparse direct solver from SciPy \cite{paper:scipy} for solving the linear systems generated by FEM. We used Numpy \cite{paper:numpy} extensively. While we accelerated the finite elements using Numba \cite{conf:numba}, we believe that much better performance can be obtained by writing the solver in C,C++,Fortran or Julia or using high-quality open source solvers like FEniCS \cite{paper:fenics} or deal.II \cite{paper:deal.ii}. Each finite element input file is 3.3MB and output file is 1.1MB and total dataset size is approximately 49GB. Simple text based JSON files were used for input and output because of their simplicity. The simulations to generate data and train the CNN were carried out on an Intel i5-11400F 2.6 Ghz processor with 6 physical and 12 logical cores and 16 GB RAM. The OS used was Windows 10 running WSL2 with Ubuntu Linux. TensorFlow 2.4 \cite{misc:tensorflow} was used for implementing the CNN. The relevant source code can be found in \cite{misc:mlscripts}.
\section{Neural networks}
\subsection{Review}
In recent years, neural networks have been applied to various applications such as image classification \cite{paper:hinton2017}, hand written digit recognition \cite{paper:kulkarni2018}, solving differential equations and symbolic integration \cite{misc:lample2019}, solving complex partial differential equations such as the Navier-Stokes equation \cite{misc:anandkumar2020}, self-driving cars \cite{misc:agnihotri2019,misc:nvidiaselfdriving2016}, chaos \cite{paper:pathak2018}, natural language processing \cite{misc:googlenlp}, face recognition \cite{conf:taigman2014} and playing board games such as chess \cite{paper:alphazero}. Several effective Machine Learning frameworks such as Google's TensorFlow \cite{misc:tensorflow}, Facebook's PyTorch \cite{incollect:pytorch}, Scikit-Learn \cite{paper:scikit-learn} are freely available. See \cite{misc:compdeep} for a complete list. We do not cover the theory of neural networks in this work. We refer the interested reader to \cite{book:aggarwal,book:goodfellow,book:chollet,misc:cs231n,misc:andrewng,misc:udemy} and references therein for detailed information about neural networks.
\subsection{Neural networks and elasticity imaging}
Given the success achieved by neural networks on the wide variety of applications cited in the previous section, it is natural to explore the application of neural networks to the inverse problem of elasticity imaging and several recent efforts \cite{paper:pateloberai2019,misc:gu2020,paper:hoeriginsana2016} have done so. In \cite{paper:pateloberai2019}, the authors use a convolutional neural network to classify specimens into elastically heterogeneous or elastically nonlinear. Their goal is to move directly from displacements to diagnosis, circumventing the solution of the elasticity imaging inverse problem. In \cite{paper:hoeriginsana2016}, the authors use a neural network to estimate strains and stress and then calculate elastic parameters. In \cite{misc:gu2020}, the authors use a neural network which predicts elasticity distributions using residual force maps to update the weights of the neural network.

In contrast, in this work we compute the shear modulus field from the displacement or strain field using a CNN\footnote{We choose CNNs as opposed to other types of neural networks because of the success they have achieved in image classification. See, for example, \cite{paper:hinton2017}.}. There are no physical constraints involved in our work. It is purely a mapping problem from the space of displacement or strain fields to the space of the shear modulus fields. The input data for our CNN is a strain or displacement field. While multiple components of strain or displacement may be used, we use only $\epsilon_{yy}$ and $u_y$ in this work. See section (\ref{sect:cnnarch}) for more details. For each input displacement or strain field, there is a known corresponding target shear modulus field. Such a pair makes a training, validation or testing example. Starting from an initial random guess of the weights, the CNN predicts shear modulus fields, compares them with the target fields to compute the loss (objective) function and its gradient with respect to the weights of the neural network. The weights are updated using the gradient in an appropriate gradient based optimization algorithm such as Adam \cite{misc:kingma2017adam} (which we use in this work). Thus the CNN learns weights for its filters and connection weights. Using this learned information, the CNN is able to predict a shear modulus field from the input data of strain or displacement fields which is seen in figure (\ref{fig:schematic_inv}).
\subsection{Need for neural networks}
{At this point, it is worth asking the questions: Are neural networks really necessary? Would a brute-force algorithm, such as algorithm (\ref{algo:brute_ei}), suffice?.\par}
%
{\centering
\begin{minipage}{.9\linewidth}
  \begin{algorithm}[H]
    \caption{\label{algo:brute_ei}A brute force algorithm for elasticity imaging.}
    \begin{algorithmic}[1]
      \State Let there be $n_{nodes}$ nodes used to discretize the shear modulus and displacement fields.
      \State Let the shear modulus at each node be allowed to take $n_{\mu}$ discrete values.
      \State \label{algo:brute_ei_3}Store displacement fields corresponding to every possible discrete shear modulus field. 
      \State \label{algo:brute_ei_4}When an unknown displacement field is encountered, find the closest displacement field from step (\ref{algo:brute_ei_3}) and output the corresponding shear modulus field as the answer.
    \end{algorithmic}    
 \end{algorithm}
\end{minipage}
\par
}
\vspace{4mm}
The problem with algorithm (\ref{algo:brute_ei}) is that the storage required in step (\ref{algo:brute_ei_3}) and the search space in step (\ref{algo:brute_ei_4}) is beyond enormous. There are ${n_{\mu}}^{n_{nodes}}$ possible shear modulus fields. Consider $n_{nodes}=1000$ and $n_{\mu}=10$, yielding $10^{1000}$ possible shear modulus fields, an enormous number. We need an algorithm to search this large search space efficiently, and hence, the need for neural networks.
\subsection{\label{sect:cnnarch} CNN architecture used in this work}
\begin{figure}[h]
  \centering
  \includegraphics[width=\linewidth]{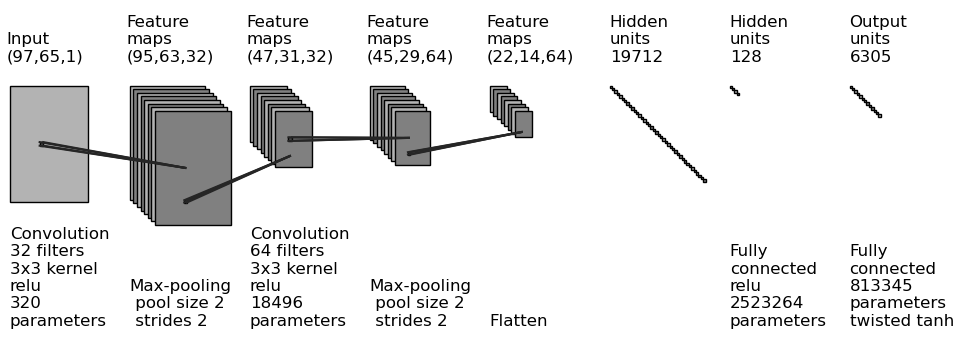}
  \caption{\label{fig:cnn_arch}CNN architecture used in this work. This figure was generated using \url{https://github.com/gwding/draw_convnet}. See also table (\ref{table:cnnparams}).}
\end{figure}

Figure (\ref{fig:cnn_arch}) shows the architecture and parameters of the CNN we use in this work. This architecture is essentially the same the one used in the deep learning example in \cite{misc:udemy} in which it was used to classify images into two categories: 'cat' or 'dog'.

The first and second dimensions in the input represent the number of nodes in the $y$ and $x$ direction respectively. The last dimension represents the number of channels in the image. In color image-processing the number of channels is typically $3$, one channel each for the three colors red, green and blue (RGB). In our case, the number of channels is the number of components of the displacement or strain field used in our problem. If we use all three components of the strain field $\epsilon_{xx},\epsilon_{yy}$ and $\epsilon_{xy}$, then the number of channels is $3$. If we use only one component of the strain field, say $\epsilon_{yy}$, then the number of channels is $1$.

Since we are working with almost incompressible linear elasticity, we know \cite{book:hugheslinear} that:
\begin{align}
  \epsilon_{xx}+\epsilon_{yy}\approx{0} \implies \pdd{u_x}{x} + \pdd{u_y}{y} \approx{0} \label{eqn:strainincomp}. 
\end{align}
In addition, since our problem setup is much closer to a compression rather than shear problem, $\epsilon_{xy}$ is close to zero almost everywhere. We therefore choose to not work with $\epsilon_{xy}$. Equation (\ref{eqn:strainincomp}) links $\epsilon_{xx}$ with $\epsilon_{yy}$ and also $u_x$ with $u_y$. Therefore, effectively, there is only one independent component of strain or displacement. In elasticity imaging using ultrasound, we typically choose to work with the displacement or strain component along the direction of ultrasound propagation because it can be estimated much more accurately than the transverse component. This also renders the shear strain $\epsilon_{xy}$ noisy and is another reason to not use shear strain data. In this work, we choose $u_y$ and $\epsilon_{yy}$ as the independent components of displacement and strain and we restrict our attention to only single channel images which represent either $u_y$ or $\epsilon_{yy}$.
\begin{table}\centering
  \ra{1.3}
  \begin{tabular}{ccc}
    \toprule
    {CNN Name} &  {Training data}  & {Prediction data}\\
    \midrule
    \multirow{3}{*}{CNN1}    &  $\epsilon_{yy}$                           &  {$\epsilon_{yy}$ + 40dB noise}\\
                              &  noiseless                                & 1-3 inclusions (similar to training)\\
                              &  1-3 inclusions                           & cross shaped example\\
     \midrule
     \multirow{2}{*}{CNN2}    &  $u_{y}$, noiseless                        & {$u_{y}$ + 40dB noise}\\
                              &  1-3 inclusions                            &  1-3 inclusions (similar to training)\\
     \midrule
     \multirow{2}{*}{CNN3}    &  $\epsilon_{yy}$, noiseless                & {$\epsilon_{yy}$ + 40dB noise}\\
                              &  1 inclusion                               & 1-3 inclusions (not similar to training)\\
    \bottomrule
  \end{tabular}
  \caption{\label{table:cnnparams} Table of CNNs trained and their parameters. All CNNs use the twisted tanh [0.75,5.25] activation function (see section (\ref{sect:outputact})). CNN1 and CNN2 have the same true shear modulus fields. All CNNs are trained on data containing circular inclusions. See also figure (\ref{fig:cnn_arch}).} 
\end{table}

We train three CNNs whose parameters are are listed in table (\ref{table:cnnparams}). There are either $1,2$ or $3$ circular inclusions in each training example used to train CNN1 and CNN2 and there is exactly one circular inclusion in each training example used to train CNN3. CNN1 and CNN2 are tested on data similar to their training data. That is, the test data also consists of a displacement or strain field corresponding to  $1,2$ or $3$ circular inclusions.  CNN1 is additionally tested on a cross shaped example which lies outside its training data, because it is not circular. We train CNN3 with data containing only $1$ inclusion but test it on data containing $1,2$ or $3$ inclusions. 

Each CNN has approximately $3.3$ million train-able parameters. We have $6000$ training images, each with $65\times97$ data points for the shear modulus. Hence, the total number of data points for the shear modulus are $6000\times97\times65=37.8$ million, approximately $11$ times the number of parameters being estimated.

The \textit{loss function} is \textit{mean squared error} and the optimizer is \textit{Adam} \cite{misc:kingma2017adam} with default TensorFlow settings. After training, we choose the model with the best validation loss make predictions. No regularization or dropout is used. We discuss output layer activations in the following section.
%
%
\subsection{\label{sect:outputact} Reduction of haloes using twisted tanh}
\begin{figure}[!h]
  \centering
  \begin{subfigure}[t]{\nhgactfuncwidth}
    \centering
    \includegraphics[totalheight=\nhgactfuncheight]{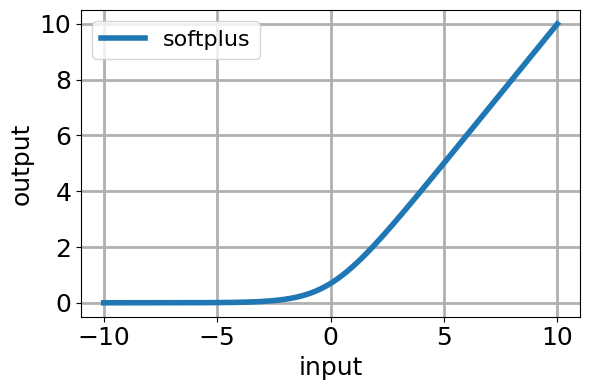}
    \subcaption{softplus}
  \end{subfigure}
  \begin{subfigure}[t]{\nhgactfuncwidth}
    \centering
    \includegraphics[totalheight=\nhgactfuncheight]{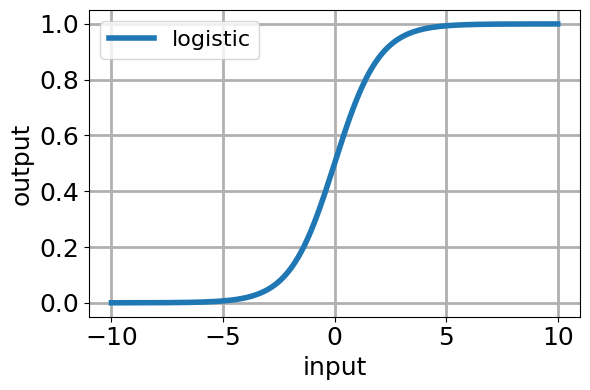}
    \subcaption{logistic}
  \end{subfigure}
  \begin{subfigure}[t]{\nhgactfuncwidth}
    \centering   
    \includegraphics[totalheight=\nhgactfuncheight]{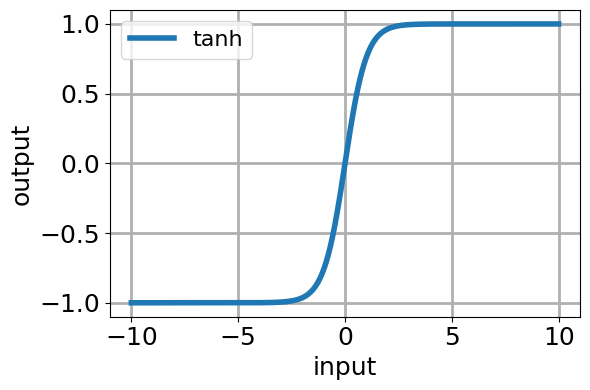}
    \subcaption{tanh}
  \end{subfigure}
  \begin{subfigure}[t]{\nhgactfuncwidth}
    \centering
    \includegraphics[totalheight=\nhgactfuncheight]{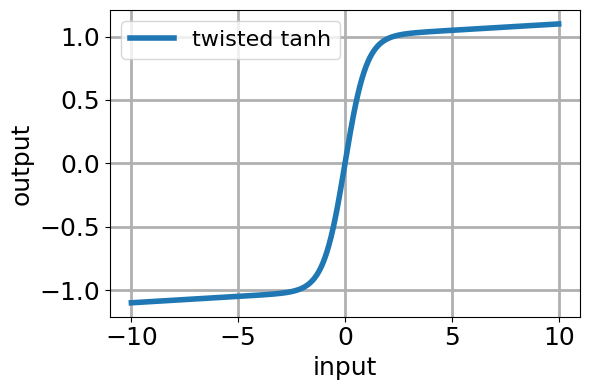}
    \subcaption{twisted tanh}
  \end{subfigure}
\caption{\label{fig:activations} Graphs of activation functions for equations (\ref{eqn:softplus}-\ref{eqn:twisttanh}).}
\end{figure}

We consider the functions given in equations (\ref{eqn:softplus}-\ref{eqn:twisttanh}) as activation functions for the output layer and choose the twisted tanh function, given in equation (\ref{eqn:twisttanh}), as the output layer activation. See figure (\ref{fig:activations}) for graphs of the activation functions. All our work, apart from the comparisons in this section, is carried out using the twisted tanh activation function.
\begin{subequations}
\begin{flalign}
&& f(x) &= \ln(1+\exp(x))   &&\text{softplus activation} \label{eqn:softplus}\\
&& f(x) &= \frac{1}{1+\exp(-x)} &&\text{logistic activation} \label{eqn:logistic}\\
&& f(x) &= \tanh(x)    &&\text{tanh activation} \label{eqn:tanh}\\
&& f(x) &= \tanh(x) + 0.01x &&\text{twisted tanh activation} \label{eqn:twisttanh}
\end{flalign}
\label{eqn:activations}
\end{subequations}

\begin{figure}[!h]
  \centering
  \begin{subfigure}[c]{\nhghaloeswidth}
  \centering    
    \includegraphics[totalheight=\nhghaloesheight]{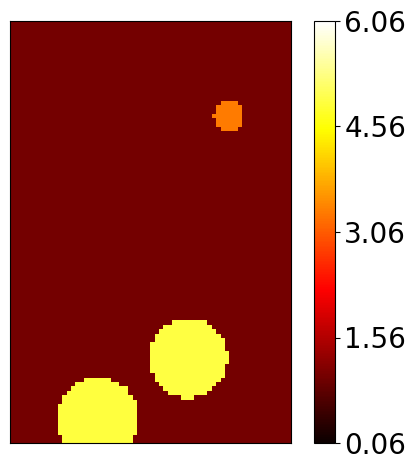}
    \caption{\label{fig:haloes_true} True}
  \end{subfigure}
  \begin{subfigure}[c]{\nhghaloeswidth}
  \centering    
    \includegraphics[totalheight=\nhghaloesheight]{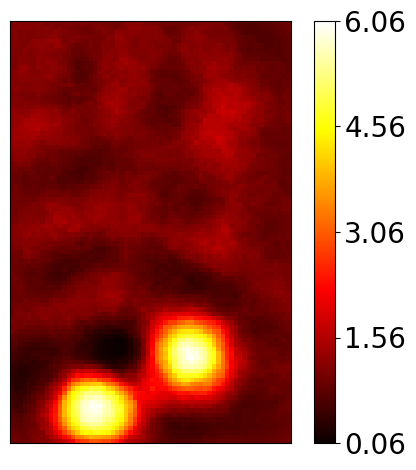}
    \caption{\label{fig:haloes_softplus} Softplus}
  \end{subfigure}
  \begin{subfigure}[c]{\nhghaloeswidth}
    \centering
    \includegraphics[totalheight=\nhghaloesheight]{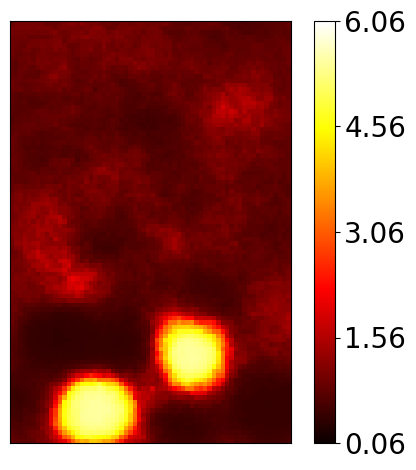}
    \caption{\label{fig:haloes_tanhp50} Twisted 1}            
  \end{subfigure}
  \begin{subfigure}[c]{\nhghaloeswidth}
    \centering
    \includegraphics[totalheight=\nhghaloesheight]{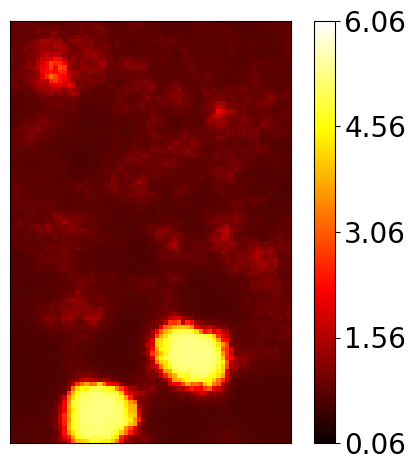}
    \caption{\label{fig:haloes_tanhp25} Twisted 2}        
  \end{subfigure}
  \begin{subfigure}[c]{\nhghaloeswidth}
    \centering
    \includegraphics[totalheight=\nhghaloesheight]{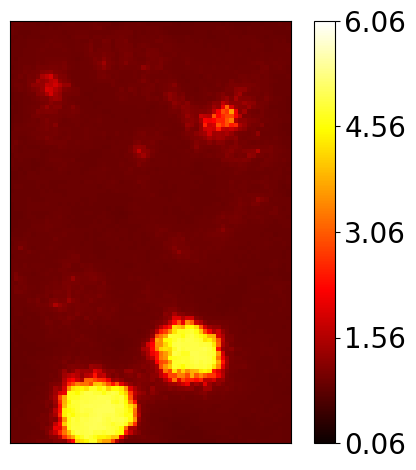}
    \caption{\label{fig:haloes_tanhp0} Twisted 3}    
  \end{subfigure}     
  \caption{\label{fig:haloes} Reconstructions using the softplus activation function show regions (black) of very low shear modulus (\textit{haloes}), typically adjoining inclusions. Twisted 1, Twisted 2 and Twisted 3 correspond to the twisted tanh $[0.5, 5.5]$, twisted tanh $[0.75,5.25]$ and twisted tanh $[1.0,5.0]$ activations. The better our prior knowledge about the shear modulus, the lesser the \textit{haloes}. These figures are on the same color scale. See also table (\ref{table:muminmax}).}
\end{figure}

The softplus activation function (\ref{eqn:softplus}) has range $(0,\infty)$ and thus it respects the physical positivity constraint on the shear modulus: $\mu(x)>0$ for reasonable materials \cite{book:segelmathcont}. The drawback of the softplus activation function is that it produces regions in which the shear modulus is very close to zero (black regions in figure (\ref{fig:haloes})). We call such regions \textit{haloes}. They are especially prominent in figures (\ref{fig:haloes_softplus}) and (\ref{fig:haloes_tanhp50}).

We now discuss how the choice of the activation function for the output layer can control \textit{haloes}. The essential idea is to bound the output shear modulus within a user-specified range. We choose the logistic and tanh activation functions, given by equations (\ref{eqn:logistic}) and (\ref{eqn:tanh}) respectively because their range is limited to $(-1,1)$ . This implies that the shear modulus fields in the training data must also be scaled such that they lie in the same interval $(-1,1)$. This requires prior knowledge of the maximum and minimum value of the shear modulus, denoted by $\mu_{lower}$ and $\mu_{upper}$, to scale the training data using a linear transformation such that the interval $[\mu_{lower},\mu_{upper}]$ is mapped linearly to the interval $[-1,1]$. We find, as noted in \cite{bookchap:lecun98b}, that the gradient of the logistic and tanh functions vanishes as we go farther away from zero. This causes the weights to get stuck and as a consequence the neural network does not learn.

To avoid this, we add, as noted in \cite{bookchap:lecun98b}, a \textit{twisting term} $0.01x$ to the the tanh activation, given in equation (\ref{eqn:tanh}), to get the twisted tanh activation function, given by equation (\ref{eqn:twisttanh}). The addition of the twisting term ensures that the gradient of the twisted tanh function does not vanish far away from zero. While the addition of the twisting term changes the range of the tanh function from $(-1,1)$ to $(-\infty,\infty)$, we see that in practice that the predictions of the neural network, when rescaled, lie only slightly outside the interval $[\mu_{lower},\mu_{upper}]$. This is seen in table (\ref{table:muminmax}) where the minimum and maximum values of the shear modulus $\mu$, over all test examples, are only slightly outside the interval $[1.0,5.0]$.

We note that, when we write: ``twisted tanh $[\mu_{lower},\mu_{upper}]$'' we mean that we are using the twisted tanh activation function with the training shear modulus data being scaled such that the interval $[\mu_{lower},\mu_{upper}]$ is mapped linearly to $[-1,1]$. 

\begin{table}
  \ra{1.3}
  \centering
  \begin{tabular}{cccc}
    \toprule
    {Activation}  & {Minimum $\mu$} & {Maximum $\mu$} & {Average scaled error}\\
    \midrule
    softplus                     &    0.0613    &  6.86       &  0.221 \\
    twisted tanh $[0.5, 5.5]$    &    0.382     &  5.54       & 0.215  \\     
    twisted tanh $[0.75,5.25]$   &    0.626     &  5.34       & 0.197  \\
    twisted tanh $[1.0, 5.0]$    &    0.805     &  5.07       & 0.191\\
    \bottomrule
  \end{tabular}
  \caption{\label{table:muminmax} Maximum and minimum values of the shear modulus (over all test examples) and the scaled error for CNNs using the softplus and twisted tanh activation functions. We see that the more accurate our prior knowledge of the shear modulus, the more accurate the predictions. See also figure (\ref{fig:haloes}). See equation (\ref{eqn:averagescalederror}) for the definition of the average scaled error.}
\end{table}
In order to compare the softplus and twisted tanh activations we define two metrics: the \textit{scaled error} and the \textit{average scaled error}. We denote the scaled error for the $i^{th}$  test example by $\zeta_{i}$ and the average scaled error by $\zeta_{ave}$. They are defined as 
  \begin{align}
  \zeta_{i} &\coloneqq \frac{\|\mu^{i,predicted} - \mu^{i,true}\|_{2}}{\|\mu^{i,true}\|_{2}} &\text{and} &&\zeta_{ave} &\coloneqq \frac{\sum_{i=1}^{n_{test}}\zeta_{i}}{n_{test}}. &\label{eqn:averagescalederror}
  \end{align}
where $\mu^{i,predicted}$ and $\mu^{i,true}$ are vectors containing the predicted and true values of the shear modulus for the $i^{th}$ test example and $n_{test}$ is the number of test examples. $\|\cdot\|_2$ is the Euclidean norm defined in equation (\ref{eqn:eucnorm}).

Referring to figure (\ref{fig:haloes}), we see that the \textit{haloes} seen in figures (\ref{fig:haloes_softplus}) and (\ref{fig:haloes_tanhp50}) are almost eliminated in figures (\ref{fig:haloes_tanhp25}) and (\ref{fig:haloes_tanhp0}). We see that the more accurately we know the true range of the shear modulus, the lesser the \textit{haloes} and closer the the maximum and minimum values of the predicted modulus, over all test examples, are to the true maximum and minimum values. From table (\ref{table:muminmax}), we also find that the average scaled error, defined above in equation (\ref{eqn:averagescalederror}), decreases as the known bounds $[\mu_{lower},\mu_{upper}]$ approach true bounds $[1.0,5.0]$. We conclude that it is possible to minimize haloes with accurate knowledge about the minimum and maximum shear modulus in the problem in the twisted tanh activation.

We close by commenting that enforcing constraints on the shear modulus is easy in traditional iterative methods because we can just supply bound constraints to the optimization algorithm. On the other hand, for neural networks, constraining the shear modulus requires constraining the range of the output layer activation which causes the gradients of the activation function to be zero, as the input goes further away from zero. This problem can be alleviated by using the twisted tanh activation.
%
%
\section{Results}
We present results for the CNNs described in table (\ref{table:cnnparams}) and section (\ref{sect:cnnarch}). Their architecture and parameters are given in figure (\ref{fig:cnn_arch}). In sections (\ref{sect:resultscnn1}) and (\ref{sect:resultscnn2}) we test the CNNs on data similar to training data. That is, both the training and test data contain displacement or strain fields corresponding to $1,2$ or $3$ circular inclusions in the domain. In sections (\ref{sect:resultscnn3}) and (\ref{sect:resultscross}) we test the performance on the CNNs on data which is unlike its training data, because the training data consists of displacement or strain fields corresponding to exactly one circular inclusion in the domain, while the test data contains either $1,2$ or $3$ circular inclusions or a non-circular, cross-shaped inclusion. 

We train each CNN for 384 epochs to put them on a equal footing for comparison. At approximately ${28}$ seconds per epoch, each network takes about ${3}$ hours to train. All CNNs use the twisted tanh $[0.75,5.25]$ activation. This means that we instruct the CNNs to predict shear moduli approximately in the range $[0.75,5.25]$ as opposed to the true range $[1.0,5.0]$. Thus, we are not assuming perfect knowledge of the upper and lower bounds for the shear modulus in the problem. For each CNN, we present two representative reconstructions which illustrate the performance of our neural network. We present histograms based on the scaled error, equation (\ref{eqn:averagescalederror}), and additional supporting reconstructions in Appendix (\ref{sect:appendix1}).

\subsection{Error metrics}
We introduce two metrics, the \textit{inclusion scaled error} and \textit{background scaled error}, to evaluate the performance of the neural networks. These metrics  measure the average percentage error in the shear modulus in the background or in the inclusion. We denote the inclusion scaled error for the $i^{th}$ test example by $\eta_{i}$ and its average by $\eta_{ave}$. They are defined as 
  \begin{align}
  \eta_{i} &\coloneqq \frac{1}{n_{inc_i}}\sum_{j=1}^{n_{inc_i}}{\Bigg (}\frac{\mu^{i,predicted}_{j}-\mu^{i,true}_{j}}{\mu^{i,true}_{j}}{\Bigg )} &\text{and}  && \eta_{ave} &\coloneqq \frac{\sum_{i=1}^{n_{test}}\eta_{i}}{n_{test}}. &\label{eqn:averageincscalederror}
  \end{align}
In the above equations, $\mu_{j}^{i,true}$ is the $j^{th}$ component of a vector $\mu^{i,true}$ containing true values of the shear modulus for the $i^{th}$ test example, \textit{restricted to the true inclusion(s)} only. $\mu_{j}^{i,predicted}$ is the $j^{th}$ component of a vector $\mu^{i,predicted}$ containing corresponding predictions. $n_{inc_i}$ is the number of nodes in the true inclusion(s) for the $i^{th}$ test example. $n_{test}$ is the number of test examples. The quantities $\eta_i$ and $\eta_{ave}$ capture how accurately the shear modulus of the inclusions is predicted. Positive values of $\eta_i$ and $\eta_{ave}$ indicate over-prediction. Negative values indicate under-prediction.

The background scaled error for the $i^{th}$ example $\xi_{i}$ and its average $\xi_{ave}$ can be defined similarly, by replacing $n_{inc_i}$ in equation (\ref{eqn:averageincscalederror})  with $n_{back_i}$ (the number of nodes in the background in the true modulus field) and restricting the vectors $\mu^{i,true}$ and $\mu^{i,predicted}$ to the background only. They are defined as 
  \begin{align}
  \xi_{i} &\coloneqq \frac{1}{n_{back_i}}\sum_{j=1}^{n_{back_i}}{\Bigg (}\frac{\mu^{i,predicted}_{j}-\mu^{i,true}_{j}}{\mu^{i,true}_{j}}{\Bigg )} &\text{and} && \xi_{ave} &\coloneqq \frac{\sum_{i=1}^{n_{test}}\xi_{i}}{n_{test}}. &\label{eqn:averagebackscalederror}
  \end{align}

%
\subsection{\label{sect:resultscnn1}Results for CNN1}
In this section, we present results for CNN1 which is trained on the axial strain in the $y$ direction, $\epsilon_{yy}$. The parameters for this CNN are given in table (\ref{table:cnnparams}) and figure (\ref{fig:cnn_arch}) and the architecture is described in section (\ref{sect:cnnarch}). This CNN is tested on data similar to its training data. This means that both training and testing data sets contain $1,2$ or $3$ circular inclusions.

\begin{figure}[!h]
  \centering
  \begin{subfigure}[c]{\nhglosswidth}
    \centering
    \includegraphics[totalheight=\nhglossheight]{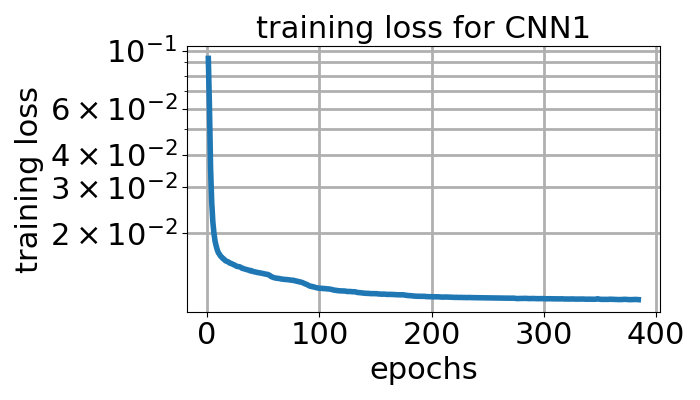}
    \subcaption{Loss}
  \end{subfigure}
  \begin{subfigure}[c]{\nhglosswidth}
    \centering    
    \includegraphics[totalheight=\nhglossheight]{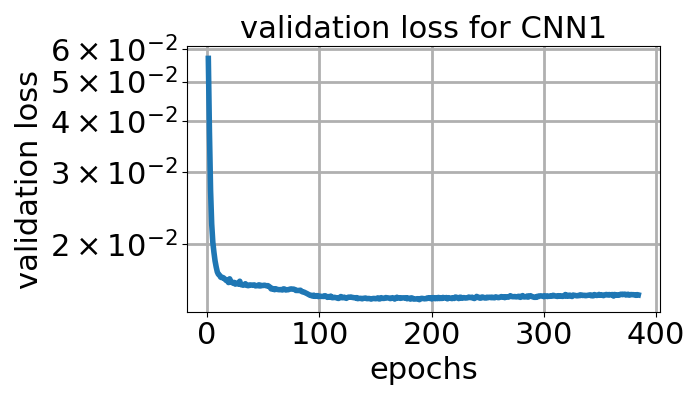}
    \subcaption{Validation loss}
  \end{subfigure}
  \caption{\label{fig:cnn1losses} Training and validation losses for CNN1.}
\end{figure}

The training and validation losses for CNN1 are shown in figure (\ref{fig:cnn1losses}). The average inclusion scaled error given in equation (\ref{eqn:averageincscalederror}) is $-0.165$ and the average background scaled error given in equation (\ref{eqn:averagebackscalederror}) is $-0.0215$. These indicate that, on average, the shear modulus of the inclusions is under-predicted. The background, on average, is only slightly under-predicted. See table (\ref{table:cnnstatsummary}) for a comparison across CNNs.

\begin{figure}[!h]
  \centering
  \begin{subfigure}[c]{\nhgcnnwidth}
    \centering
    \includegraphics[totalheight=\nhgcnnheight]{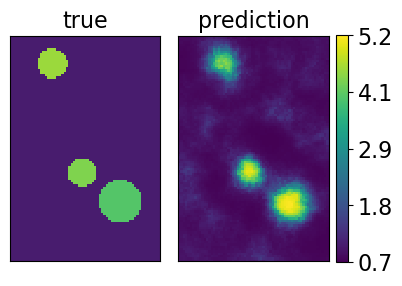}
    \subcaption{\label{fig:cnn1resulta} (0.298)}
  \end{subfigure}
  \begin{subfigure}[c]{\nhgcnnwidth}
    \centering
    \includegraphics[totalheight=\nhgcnnheight]{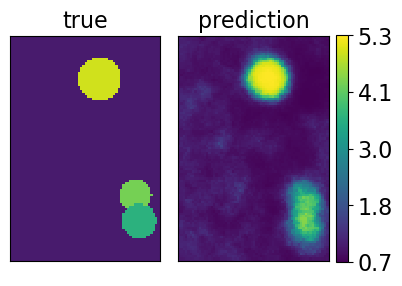}
    \subcaption{\label{fig:cnn1resultb} (0.247)}
  \end{subfigure}
\caption{\label{fig:cnn1result} Representative results for CNN1. The numbers in the brackets are the scaled errors as defined in equation (\ref{eqn:averagescalederror}).}  
\end{figure}

Representative reconstructions are shown in figure (\ref{fig:cnn1result}). The location of the inclusions is predicted accurately while the shear modulus predictions have small errors. The shear modulus of inclusions of small size and low shear modulus is under-predicted and their geometry is predicted only in an approximate manner. We call such approximate predictions of geometry as \textit{wisps}. This effect is prominently seen in the inclusion on the top-left in figure (\ref{fig:cnn1resulta}) and the inclusion on the bottom right in figure (\ref{fig:cnn1resultb}). We present additional supporting results in figure (\ref{fig:app1result}) in Appendix (\ref{sect:appendix1}) which lead us to the same conclusions.

\begin{figure}[!h]
\captionsetup[subfigure]{justification=centering}
  \centering
  \begin{subfigure}[c]{\nhghistowidth}
    \centering
    \includegraphics[totalheight=\nhghistoheight]{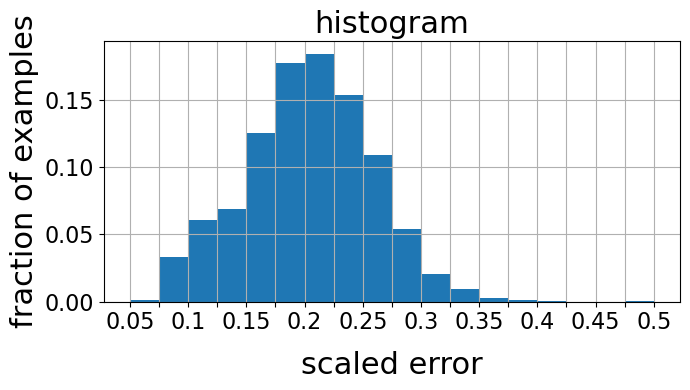}
    \subcaption{\label{fig:cnn1histoa}Histogram}
  \end{subfigure}
  \begin{subfigure}[c]{\nhghistowidth}
    \centering
    \includegraphics[totalheight=\nhghistoheight]{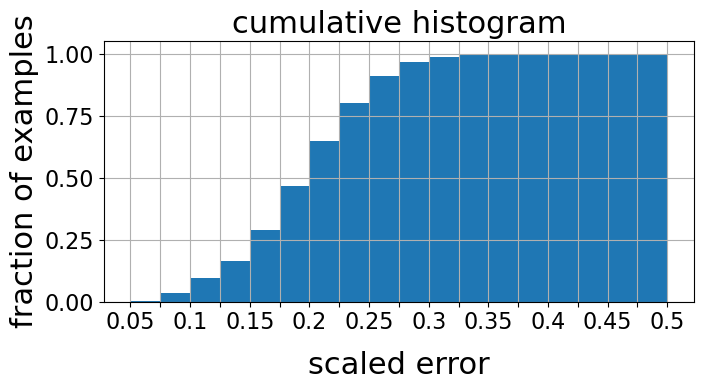}
    \subcaption{\label{fig:cnn1histob}Cumulative histogram}
  \end{subfigure}
  \caption{\label{fig:cnn1histo} Histogram and cumulative histogram for CNN1. The scaled error is defined in equation (\ref{eqn:averagescalederror}).}
\end{figure}

Figure (\ref{fig:cnn1histo}) shows histograms in which the scaled error, given in equation (\ref{eqn:averagescalederror}), is on the x-axis and the y-axis represents the fraction of examples in each bin (number of examples in each bin divided by the number of test examples). We see that the examples appear to be approximately normally distributed. The mean is $0.203$, the standard deviation is $0.0550$ and $80\%$ of the test examples have scaled error of less than $0.25$. The maximum and minimum shear moduli in the prediction, overall test examples, are $5.33$ and $0.599$ as opposed to the true values of $5.0$ and $1.0$. 
%
%
\subsection{\label{sect:resultscnn2}Results for CNN2}
In this section we present results for CNN2 which is trained on the displacement field in the $y$ direction, $u_y$. The parameters for this CNN are given in table (\ref{table:cnnparams}) and figure (\ref{fig:cnn_arch}) and the architecture is described in section (\ref{sect:cnnarch}). This CNN is tested on data similar to its training data. This means that both training and testing data sets contain $1,2$ or $3$ circular inclusions.

\begin{figure}[!h]
  \centering
  \begin{subfigure}[c]{\nhglosswidth}
    \centering
    \includegraphics[totalheight=\nhglossheight]{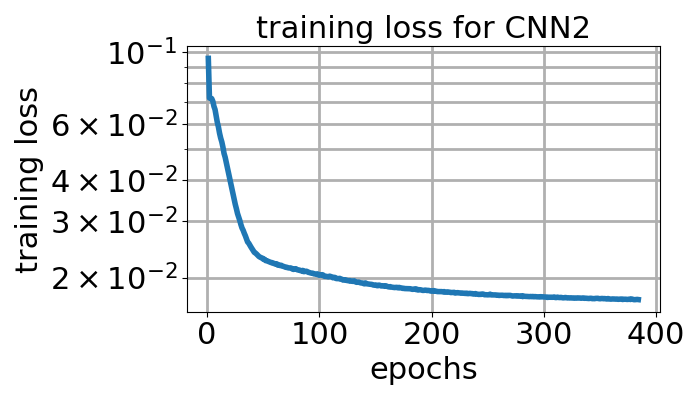}
    \subcaption{Loss}
  \end{subfigure}
  \begin{subfigure}[c]{\nhglosswidth}
    \centering
    \includegraphics[totalheight=\nhglossheight]{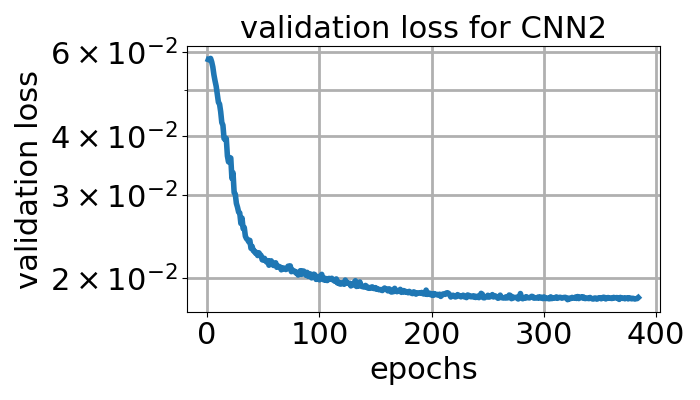}
    \subcaption{Validation loss}
  \end{subfigure}
  \caption{\label{fig:cnn2losses} Training and validation losses for CNN2.}
\end{figure}

The training and validation losses for CNN1 are shown in figure (\ref{fig:cnn2losses}). The average inclusion scaled error is $-0.309$ and the average background scaled error is $-0.107$. These indicate that, on average, the shear modulus of the inclusions is under-predicted much more than CNN1. The background is also under-predicted to a larger degree than CNN1. This indicates that in the type of CNNs studied in this work, using displacements, as opposed to strains, causes more under-prediction of the shear modulus. See table (\ref{table:cnnstatsummary}) for a comparison across CNNs.

\begin{figure}[!h]
  \centering
  \begin{subfigure}[c]{\nhghalfwidth}
    \centering
    \includegraphics[totalheight=\nhgtotalheight]{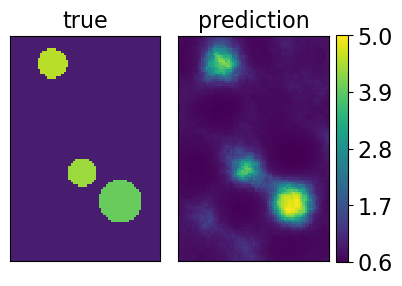}
    \subcaption{\label{fig:cnn2resulta} (0.327)}
  \end{subfigure}
  \begin{subfigure}[c]{\nhghalfwidth}
    \centering
    \includegraphics[totalheight=\nhgtotalheight]{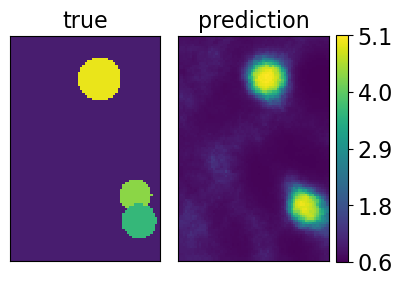}
    \subcaption{\label{fig:cnn2resultb} (0.346)}
  \end{subfigure}
\caption{\label{fig:cnn2result} Representative results for CNN2. The numbers in the brackets are the scaled errors as defined in equation (\ref{eqn:averagescalederror}). These are the same test examples presented in figures (\ref{fig:cnn1result}).}  
\end{figure}

We present representative reconstructions in figure (\ref{fig:cnn2result}). These reconstructions are for the same true shear modulus fields for CNN1 shown in figure (\ref{fig:cnn1result}). We have only changed the training and prediction data from $\epsilon_{yy}$ to $u_{y}$ and retrained the network. The scaled errors in the reconstructions in figure (\ref{fig:cnn2result}) show that the results using CNN1 are significantly more accurate than CNN2. Visually, one can see that the central inclusion in (\ref{fig:cnn1resulta}) is not clearly captured in figure (\ref{fig:cnn2resulta}). The shear modulus of the inclusions on the lower right is predicted more accurately in figure (\ref{fig:cnn1resultb}) than figure (\ref{fig:cnn2resultb}). In general, we see that the location of the inclusions is predicted accurately but the shear modulus predictions have small errors. The shear modulus of small inclusions and inclusions of low shear modulus is under-predicted more severely. This is clearly seen for the central inclusion in figure (\ref{fig:cnn2resulta}). The geometry of inclusions on the top-left and at the center of figure (\ref{fig:cnn2resulta}) is predicted as \textit{wisps}. We see in figure (\ref{fig:cnn2resultb}) that two small inclusions of relatively low shear modulus have been combined into one small inclusion of higher shear modulus. We present additional results illustrating the performance of the network in figure (\ref{fig:app2result}) in Appendix (\ref{sect:appendix1}) where we see similar results.

\begin{figure}[!h]
\captionsetup[subfigure]{justification=centering}
  \centering
  \begin{subfigure}[c]{\nhghistowidth}
    \centering
    \includegraphics[totalheight=\nhghistoheight]{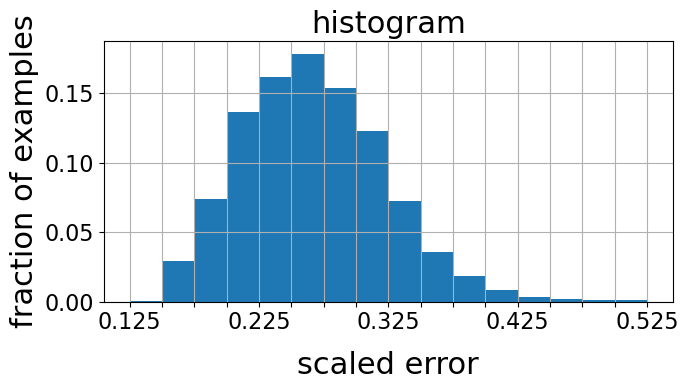}
    \subcaption{\label{fig:cnn2histoa}Histogram}
  \end{subfigure}
  \begin{subfigure}[c]{\nhghistowidth}
    \centering
    \includegraphics[totalheight=\nhghistoheight]{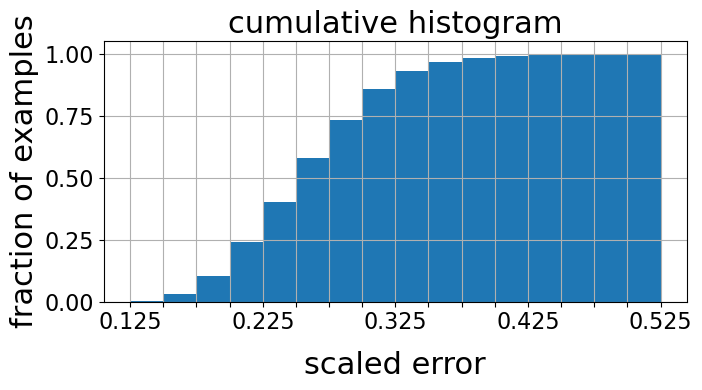}
    \subcaption{\label{fig:cnn2histob}Cumulative histogram}
  \end{subfigure}
  \caption{\label{fig:cnn2histo} Histogram and cumulative histogram for CNN2. The scaled error is defined in equation (\ref{eqn:averagescalederror}).}
\end{figure}

Figure (\ref{fig:cnn2histo}) shows histograms of the scaled error, equation (\ref{eqn:averagescalederror}). We see that the examples appear to be approximately normally distributed. The mean is $0.267$ the standard deviation is $0.0551$ and only $40\%$ test examples have scaled error of less than $0.25$. These numbers are significantly worse than the numbers for CNN1, indicating that the strain based CNN1 performs better than displacement based CNN2. The reason for this may lie in the filters learned by the two CNNs. We think that since strains are related to displacements by differentiation, an additional convolutional layer may give the displacement based CNN2 an opportunity to learn differentiating filters and improve its performance. The maximum and minimum shear moduli in the prediction, over all test examples, are $5.28$ and $0.583$ as opposed to the true values of $5.0$ and $1.0$. 
%
%
\subsection{\label{sect:resultscnn3}Results for CNN3}
\begin{figure}[!h]
  \centering
  \begin{subfigure}[c]{\nhglosswidth}
    \centering
    \includegraphics[totalheight=\nhglossheight]{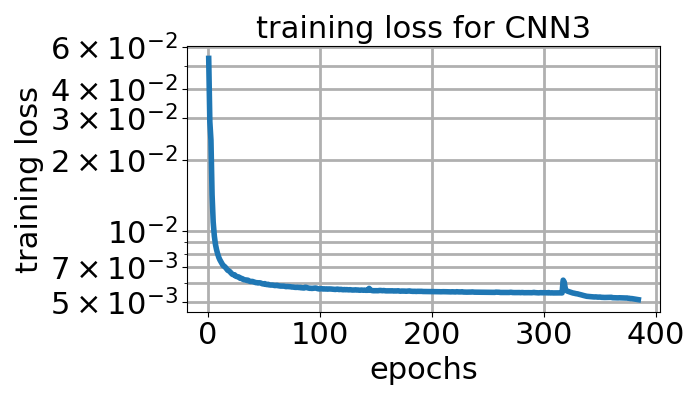}
    \subcaption{Loss}
  \end{subfigure}
  \begin{subfigure}[c]{\nhglosswidth}
    \centering
    \includegraphics[totalheight=\nhglossheight]{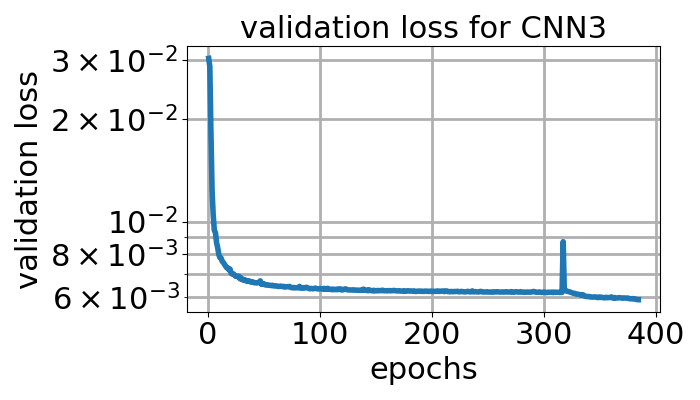}
   \subcaption{Validation loss}
  \end{subfigure}
  \caption{\label{fig:cnn3losses} Training and validation losses for CNN3.}
\end{figure}
%
\begin{figure}[!h]
  \centering
  \begin{subfigure}[c]{\nhghalfwidth}
    \centering
    \includegraphics[totalheight=\nhgtotalheight]{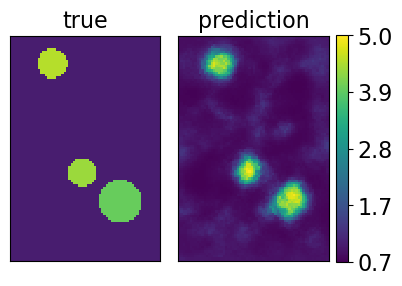}
    \subcaption{\label{fig:cnn3resulta} (0.320)}
  \end{subfigure}
  \begin{subfigure}[c]{\nhghalfwidth}
    \centering
    \includegraphics[totalheight=\nhgtotalheight]{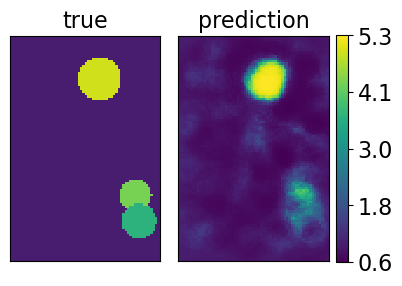}
    \subcaption{\label{fig:cnn3resultb} (0.331)}
  \end{subfigure}
\caption{\label{fig:cnn3result} Representative results for CNN3. The numbers in the brackets are the scaled errors as defined in equation (\ref{eqn:averagescalederror}). These are the same test examples presented in figures (\ref{fig:cnn1result}) and (\ref{fig:cnn2result}).}  
\end{figure}
%
\begin{figure}[!h]
\captionsetup[subfigure]{justification=centering}
  \centering
  \begin{subfigure}[c]{\nhghalfwidth}
    \centering
    \includegraphics[totalheight=\nhgtotalheight]{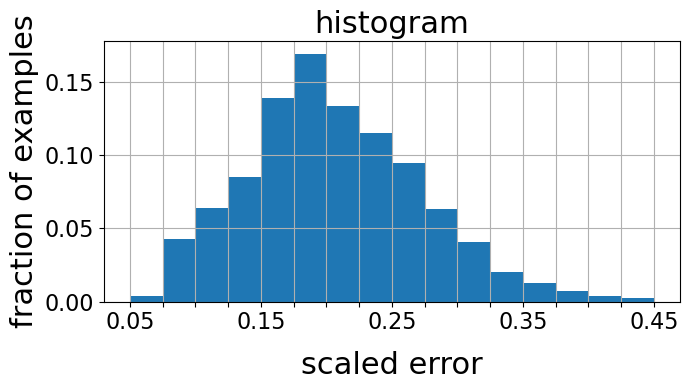}
    \subcaption{\label{fig:cnn3histoa}Histogram}
  \end{subfigure}
  \begin{subfigure}[c]{\nhghalfwidth}
      \centering
    \includegraphics[totalheight=\nhgtotalheight]{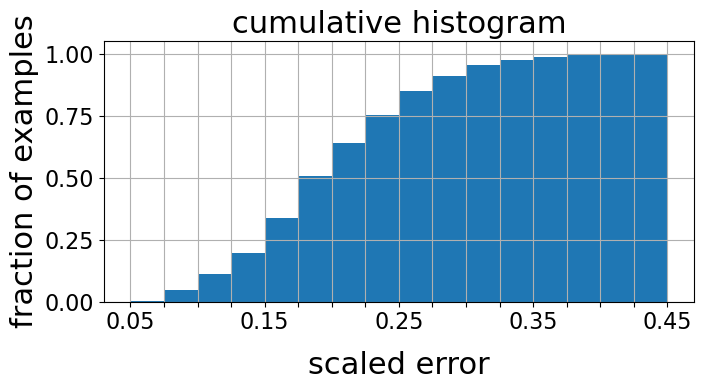}
    \subcaption{\label{fig:cnn3histob}Cumulative histogram}
  \end{subfigure}
  \caption{\label{fig:cnn3histo} Histogram and cumulative histogram for CNN3. The scaled error is defined in equation (\ref{eqn:averagescalederror}).}
\end{figure}
In this section, we evaluate the ability of the CNN to generalize to examples outside its training set. The parameters for this CNN are given in table (\ref{table:cnnparams}) and figure (\ref{fig:cnn_arch}) and the architecture is described in section (\ref{sect:cnnarch}). This CNN is trained using data contains only one circular inclusion. But we evaluate it on data containing $1,2$ or $3$ circular inclusions\footnote{This is the same data was used to train and test CNN1 and CNN2.}. We see evidence of some ability to generalize.

The training and validation losses for CNN3 are shown in figure (\ref{fig:cnn3losses}). The average inclusion scaled error is $-0.191$ and the average background scaled error is $-0.0459$. As expected, the average inclusion scaled error is worse than CNN1, but is surprisingly better than CNN2 because, unlike CNN3, CNN2 was tested with data similar to its training data, in the sense that both training and testing data contained exactly $1,2$ or $3$ circular inclusions. The average background scaled error is also worse than CNN1 but better than CNN2. This is another indication that strain based CNNs studied in this work outperform displacement based CNNs.

We present representative reconstructions in figures (\ref{fig:cnn3resulta}) and (\ref{fig:cnn3resultb}). These reconstructions are for the same true shear modulus fields for CNN1 and CNN2 (see figures (\ref{fig:cnn1result}) and (\ref{fig:cnn2result})). These results are worse than those obtained by CNN1 (figure \ref{fig:cnn1result}) and slightly better than those obtained by CNN2, figure (\ref{fig:cnn2result}), as measured by the scaled error, equation (\ref{eqn:averagescalederror}). It is interesting that a CNN trained on examples containing only one circular inclusion in the domain can detect multiple circular inclusions. Visually it appears that the performance of CNN3 in the first example, figure (\ref{fig:cnn3resulta}), is better than CNN2 but worse than CNN1. It would appear that the performance is worse than CNN1 but better than CNN2. The inclusions on the bottom right in figure (\ref{fig:cnn3resultb}) have low shear modulus and are detected only as \textit{wisps}. Visually, this result is worse than CNN1 and CNN2. We have also observed this phenomenon in other reconstructions. We present additional results illustrating the performance of the network in figure (\ref{fig:app3result}) in Appendix (\ref{sect:appendix1}).

Figure (\ref{fig:cnn3histo}) shows histograms of the scaled error. We see that the examples appear to be approximately normally distributed. The mean is $0.206$ which is only slightly worse than CNN1 ($0.203$) and much better than CNN2 ($0.267$). The standard deviation is $0.0664$ which is higher than CNN1 ($0.0550$) and CNN2 ($0.0551$). $75\%$ examples have a scaled error less that $0.25$. This similar to the corresponding number for CNN1 ($80\%$) and much better than CNN2 ($40\%$) indicating that CNNs trained on one inclusion appear to generalize well to detecting multiple inclusions. This result suggests that it may not be necessary to train CNNs with data containing multiple inclusions. The maximum and minimum shear moduli in the prediction, over all test examples, are $5.36$ and $0.524$ as opposed to the true values of $5.0$ and $1.0$.
\subsection{\label{sect:resultscross} Cross shaped example}
\begin{figure}[!h]
  \centering
  \includegraphics[totalheight=\nhgtotalheight]{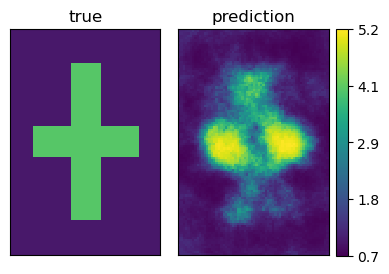}
  \caption{\label{fig:resultcross} Result for cross shaped inclusion using CNN1. True shear modulus (left), prediction (right). The scaled error, equation (\ref{eqn:averagescalederror}), is $0.418$.}
\end{figure}
Figure (\ref{fig:resultcross}) shows a reconstruction obtained when CNN1 is presented with data that is unlike its training data. We recall that CNN1 was trained using only $1,2$ or $3$ circular inclusions, but now, we are testing it with a rectangular cross shaped inclusion.

The data is the noiseless axial strain $\epsilon_{yy}$ corresponding to the true shear modulus field in figure (\ref{fig:resultcross}). The boundary conditions were the same as described before in section (\ref{sect:probsetup}). The reconstruction in figure (\ref{fig:resultcross}) shows that the CNN1 can predict that there is an inclusion but cannot detect its exact shape. This result shows that if CNNs studied in this work are presented with data unlike their training data, good results may not be obtained. This is unlike iterative methods which have been shown in \cite{paper:oberaipmb2004} to be able to capture inclusions with complicated shapes.
\subsection{\label{sect:discussion}Discussion}
\begin{table}\centering
  \ra{1.3}
  \begin{tabular}{c|ccc}
    \toprule
                              & Average      & Average                & Average\\
            CNN name          & scaled error & inclusion scaled error & background scaled error\\
    \midrule
    CNN1                      & $0.203$   & $-0.165$     & $-0.0215$     \\
    CNN2                      & $0.267$   & $-0.309$     & $-0.107$      \\
    CNN3                      & $0.206$   & $-0.191$     & $-0.0459$     \\
  \bottomrule  
  \end{tabular}
  \caption{\label{table:cnnstatsummary} Errors for the CNNs considered. See equation (\ref{eqn:averagescalederror}) for the average scaled error, equation (\ref{eqn:averageincscalederror}) for the average inclusion scaled error, and equation (\ref{eqn:averagebackscalederror}) for the average background scaled error.}
\end{table}
We observe encouraging results with good agreement between the true and predicted shear shear modulus images. Table (\ref{table:cnnstatsummary}) summarizes the performance of the three CNNs evaluated in this work\footnote{We may be able to use these numbers to create fudge-factors increase the shear modulus of the inclusions predicted by the neural network.}. The shear modulus of inclusions is under-predicted. The shear modulus of inclusions with low shear modulus and small size is significantly under-predicted, resulting in predictions that can be best described as \textit{wisps}.

The average inclusion scaled error, given in equation (\ref{eqn:averageincscalederror}) evaluated over inclusions having shear modulus in the range $[2.0,3.0)$ yields $-0.206$,$-0.347$ and $-0.247$ for CNN1, CNN2 and CNN3 respectively. This shows that all three networks perform worse than usual while detecting inclusions which whose shear modulus is low. 
On the basis of the average scaled error, table (\ref{table:cnnstatsummary}),  CNN1 outperforms CNN3 which in turn outperforms CNN2. From cumulative histograms shown in figures (\ref{fig:cnn1histob}),(\ref{fig:cnn2histob}) and (\ref{fig:cnn3histob}), the fraction of examples classified with average scaled error less than $0.25$ was $0.80$ for CNN1 and $0.40$ for CNN2 and $0.75$ for CNN3. Thus, the strain based networks CNN1 and CNN2 are more accurate at predicting the value of the shear modulus of the inclusion than the displacement based network CNN3. We think that exploring the physical meaning of the convolutional filters learned using Fourier analysis as in \cite{paper:pateloberai2019} is a promising direction for investigating the reason for out-performance.

The results presented in section (\ref{sect:resultscnn3}) show that a CNN trained on one inclusion in the domain can generalize to multiple inclusions in the domain. The result in section (\ref{sect:resultscross}) shows that CNN1, trained on circular inclusions, does not generalize perfectly to non-circular examples. 
\section{Concluding remarks}
In this work, we have presented CNNs capable of predicting shear modulus fields from displacement or strain fields and obtained results that warrant further research. We use the twisted tanh activation function to constrain the shear modulus prediction as per our prior knowledge and reduce \textit{haloes}. The shear modulus of the inclusions with shear modulus close to the background modulus, or small inclusions, is predicted as \textit{wisps}, while regions of high shear modulus are predicted more accurately. It is seen that while the CNNs exhibit good performance on the type of examples whose geometry is similar to the examples on which they were trained. They have limited ability to generalize to examples with unseen geometry, as is seen with the cross-shaped example. Using bigger data sets containing a large number of examples with different geometric characteristics may alleviate this problem.
\subsection{Directions for future work}
\begin{enumerate}
\item{We can consider expanding the datasets for training the CNNs by adding homogeneous examples. We expect good performance on such datasets because CNNs under-predict the shear modulus of small inclusions. Another direction would be to first classify the test or prediction example under consideration as homogeneous or non-homogeneous and then reconstruct the shear modulus field if it is deemed to be non-homogeneous.}
\item{We think that training the CNN with displacement or strain data generated using random values for the shear modulus at each node, or perturbing the shear modulus of only one node relative to the background, and seeing if it learns the inverse operator from the displacement (or strain) fields to shear modulus fields will be interesting.}
\item{Stiffness of smaller inclusions and inclusions with low shear modulus is under-predicted. We note that from previous experience, the adjoint method \cite{paper:oberai2003,paper:oberaipmb2004,paper:gokhale2008} is able to predict the small inclusions correctly using noiseless data. Designing a network to accurately image small inclusions and inclusions of low shear modulus will be interesting. We note that, this this work,  material properties and displacements were on the same mesh. This may lead to unconverged displacements for smaller inclusions which may not have enough information to be of use in predicting small inclusions.}
\item{We think that Fourier analysis of these filters, as reported in \cite{paper:pateloberai2019} would be valuable. Using their technique, we can identify some filters in this work as the derivative in the $x$ or $y$ direction, but the physical meaning of the majority of filters is not clear. We believe that after understanding the physical meaning of each filter, better filters could perhaps be constructed manually. Visualization of intermediate images produced by the convolutional layers is also a promising direction. We think that seeking visual explanations \cite{paper:selvaraju_2019} for the CNNs considered in this work will also be interesting.}
\item{We used a simple \textit{mean squared error} (which corresponds to the square of the $L^2$ norm in the continuous case) as the loss function for the neural network. The effect of other losses corresponding to $H^1$ norm or the $L^2$ norm with the addition of the Total Variation Diminishing (TVD) norm will be interesting to evaluate. We think that, based on our experience in \cite{diss:gokhale2007}, using a loss function corresponding to the $H^1$ norm will remove the incidence of regions of low shear modulus adjacent to inclusions. However, it will also penalize sharp discontinuities in the inclusions as well. If we are training with both components of a noiseless displacement field, then one can consider a loss function which is simply an appropriate norm of the  residual obtained when the predicted shear modulus and input fields are inserted into the equations of elasticity. This would result in a network similar to the \textit{physics-informed neural networks} discussed in \cite{paper:pinnkarniadakis}}
\item{Investigating different network architectures such as ResNet \cite{conf:resnet} or VGG \cite{conf:vgg} with the aim of yielding better results will be worth investigating. More CNN layers (or less), deeper (or shallower) networks, different kernel sizes for convolutional layers should be investigated. The need for \textit{max pooling} layers should particularly be investigated because they reduce the size of the input image and may thus cause a loss of resolution. It may be worth investigating whether a CNN is required at all and whether a simple dense network can produce similar quality results. Another direction would be to change the number of nodes in the first dense layer which contains $128$ nodes. Increasing this number will probably result in better networks, but will also increase the number of weights and hence training time. It may also be worthwhile to replace the full connection of the dense layers with spatially close connections. By this, we mean that after flattening each node in the first dense layer would be connected only to those neurons which are spatially close together in the previous layer.}
\item{Medical image registration to obtain a displacement field is a difficult process. It would be an important advance if we could train neural networks to work directly with medical images instead of the computed displacement field. This would involve training the CNN by computing thousands of displacement fields by hand and solving an inverse problem and using the predicted shear modulus field as labeled data as input to the CNN. Additional information could be obtained by doctors interpreting medical images and identifying tumors and their mechanical properties. A first step in this direction would be to generate artificial ultrasound images as in \cite{paper:doyley}.}
\item{We think that it would be interesting to consider a multi-scale/hierarchical neural network. This neural network would first make a prediction of the average shear modulus field using only a few weights and neurons. In the next step, more neurons would be introduced to make a prediction of the shear modulus field. At every step, the number of layers and neurons would be increased and the weights would be initialized from the previous neural network. This process can continue until a neural network which can make detailed predictions of the stiffness field can be obtained.}
\item{Extending this work to complex three dimensional organ geometries (e.g. a 3D breast geometry) discretized using unstructured finite element meshes will be interesting because the domain will no longer be rectangular and will change from patient to patient. Data will no longer be available at pre-determined spatial locations and each organ model will have different geometry.  A method that could be considered is to approximate the organ geometry using a structured grid. If the center of the cell lies inside the organ, then that entire cell is considered to be inside the organ. Cells outside the organ can have placeholder data, while cells inside the organ can have actual displacement or strain data. Because the geometry is now represented on a structured grid, simple CNNs can be used. We believe that while this method may be feasible in two dimensions, it will result in a waste of storage space in three dimensions.}
\item{Assuming that inclusions are roughly circular, we can consider several other problems. Given a displacement, strain or shear modulus field (computed by any inversion procedure) we may train a CNN to compute : 1) whether or not inclusions exist in the domain. 2) the number of inclusions in the problem 3) the shear modulus of each inclusion 4) the location of the center of each inclusion. 5) the radius of each inclusion.}
\item{A hybrid approach combining machine learning and traditional approaches (iterative or direct) will be interesting. We envision that the CNN based approach will provide a measure of its confidence in its output. If this is confidence is low, then a solution will be computed using traditional direct or iterative methods and added to the dataset.}
\item{Extending the method studied in this document to predict multiple parameter fields will be interesting, because the method will have to examine the features in the displacement or strain fields and then decide which parameter field was responsible for producing those features. Such cases will occur when we try to predict both $\lambda(x)$ and $\mu(x)$ for linear elasticity both the nonlinear parameter $\gamma(x)$ and the linear parameter $\mu(x)$ as in \cite{paper:gokhale2008}. We think separate CNNs for each material parameter will be required. We will also need to use multiple displacement fields \cite{paper:barbonegokhale,paper:barbonebamber} by increasing the number of channels in our input data. Use of multiple displacement fields will also help to predict the shear modulus field uniquely.}
\item{We think that using millions of training examples will improve the performance of the CNNs evaluated in this work. Using a large number of examples will also help incorporate large ranges of $\mu$, complex inclusion shapes, different boundary conditions, and training with many noisy displacement or strain fields generated from noiseless fields. These noisy fields, corresponding to different realizations of noise, will correspond to the same true shear modulus field. Highly optimized finite element solvers and state of the art computer hardware such as high end clusters of CPUs and GPUs and fast SSD storage will be required for data generation and CNN training.}
\end{enumerate}
\clearpage
\newpage
\begin{appendices}
\section{\label{sect:appendix1} Supporting results}
\begin{figure}[!h]
  \centering
  \begin{subfigure}[c]{\nhgappwidth}
    \centering    
    \includegraphics[totalheight=\nhgappheight]{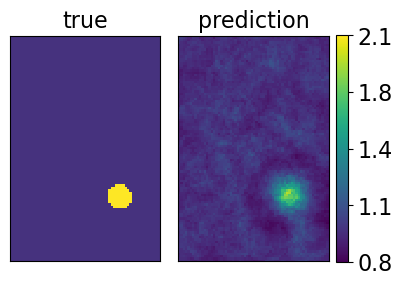}
    \subcaption{\label{fig:app1resulta} (0.0839)}
  \end{subfigure}
  \begin{subfigure}[c]{\nhgappwidth}
    \centering    
    \includegraphics[totalheight=\nhgappheight]{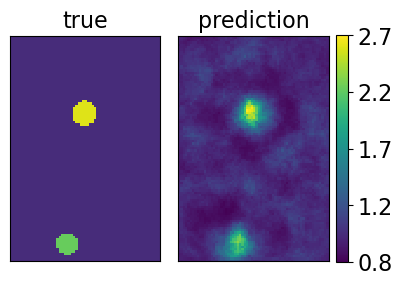}
    \subcaption{\label{fig:app1resultb} (0.138)}
  \end{subfigure}
  \begin{subfigure}[c]{\nhgappwidth}
    \centering
    \includegraphics[totalheight=\nhgappheight]{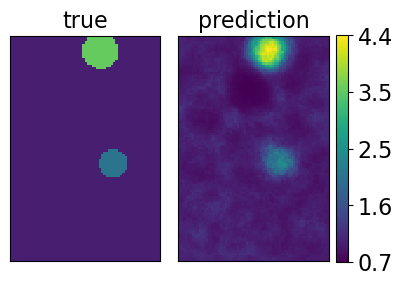}
    \subcaption{\label{fig:app1resultc} (0.168)}
  \end{subfigure}
  \begin{subfigure}[c]{\nhgappwidth}
    \centering
    \includegraphics[totalheight=\nhgappheight]{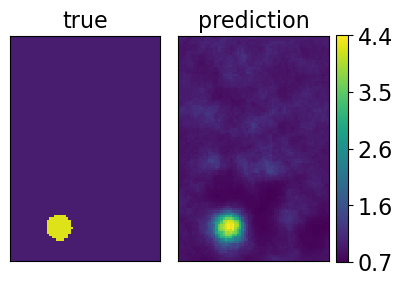}
    \subcaption{\label{fig:app1resultd} (0.185)}
  \end{subfigure}
    \begin{subfigure}[c]{\nhgappwidth}
    \centering    
    \includegraphics[totalheight=\nhgappheight]{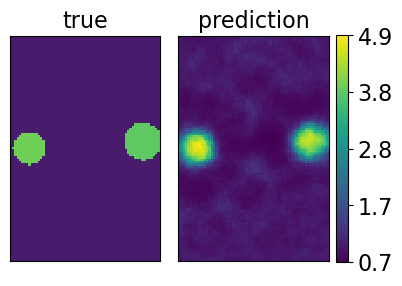}
    \subcaption{\label{fig:app1resulte} (0.205)}
  \end{subfigure}
  \begin{subfigure}[c]{\nhgappwidth}
    \centering    
    \includegraphics[totalheight=\nhgappheight]{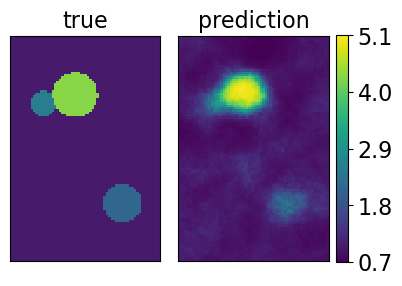}
    \subcaption{\label{fig:app1resultf} (0.222)}
  \end{subfigure}
  \begin{subfigure}[c]{\nhgappwidth}
    \centering
    \includegraphics[totalheight=\nhgappheight]{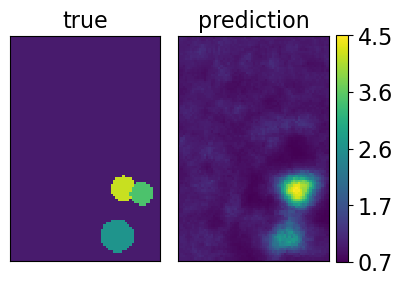}
    \subcaption{\label{fig:app1resultg} (0.243)}
  \end{subfigure}
  \begin{subfigure}[c]{\nhgappwidth}
    \centering
    \includegraphics[totalheight=\nhgappheight]{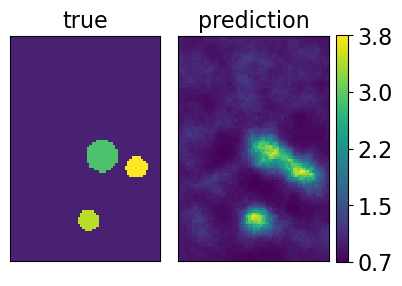}
    \subcaption{\label{fig:app1resulth} (0.264)}
  \end{subfigure}
\caption{\label{fig:app1result} Additional results for CNN1 in increasing order of the scaled error (in brackets) as defined in equation (\ref{eqn:averagescalederror}). True results are on the left and predictions on the right.}  
\end{figure}

\begin{figure}[!h]
  \centering
  \begin{subfigure}[c]{\nhgappwidth}
    \centering    
    \includegraphics[totalheight=\nhgappheight]{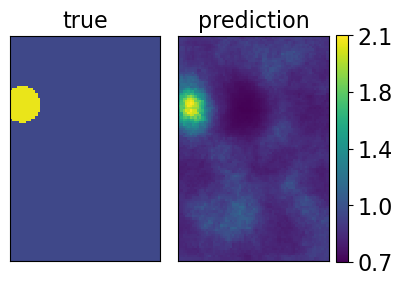}
    \subcaption{\label{fig:app2resulta} (0.156)}
  \end{subfigure}
  \begin{subfigure}[c]{\nhgappwidth}
    \centering    
    \includegraphics[totalheight=\nhgappheight]{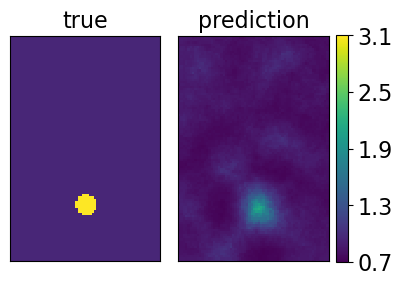}
    \subcaption{\label{fig:app2resultb} (0.205)}
  \end{subfigure}
  \begin{subfigure}[c]{\nhgappwidth}
    \centering
    \includegraphics[totalheight=\nhgappheight]{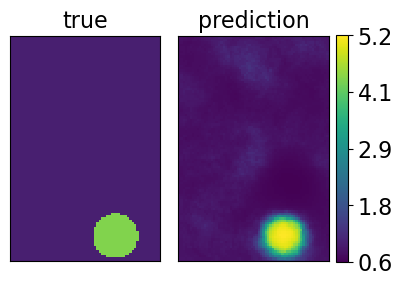}
    \subcaption{\label{fig:app2resultc} (0.227) }
  \end{subfigure}
  \begin{subfigure}[c]{\nhgappwidth}
    \centering
    \includegraphics[totalheight=\nhgappheight]{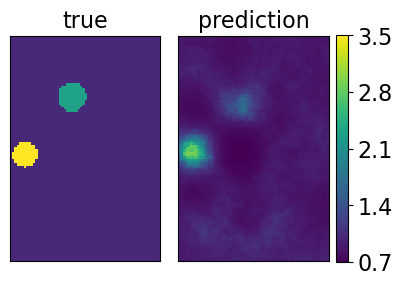}
    \subcaption{\label{fig:app2resultd} (0.247)}
  \end{subfigure}
    \begin{subfigure}[c]{\nhgappwidth}
    \centering    
    \includegraphics[totalheight=\nhgappheight]{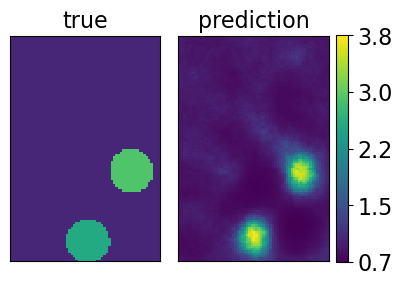}
    \subcaption{\label{fig:app2resulte} (0.265)}
  \end{subfigure}
  \begin{subfigure}[c]{\nhgappwidth}
    \centering    
    \includegraphics[totalheight=\nhgappheight]{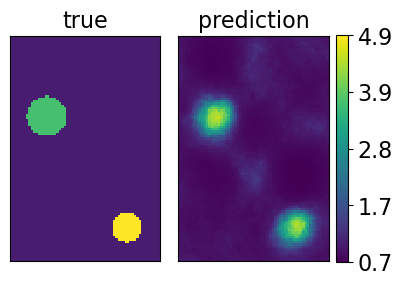}
    \subcaption{\label{fig:app2resultf} (0.283)}
  \end{subfigure}
  \begin{subfigure}[c]{\nhgappwidth}
    \centering
    \includegraphics[totalheight=\nhgappheight]{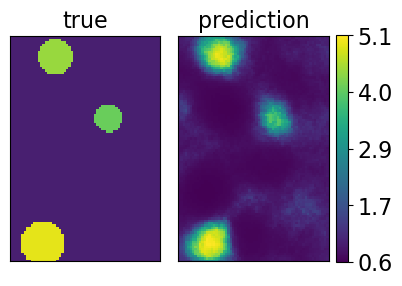}
    \subcaption{\label{fig:app2resultg} (0.312)}
  \end{subfigure}
  \begin{subfigure}[c]{\nhgappwidth}
    \centering
    \includegraphics[totalheight=\nhgappheight]{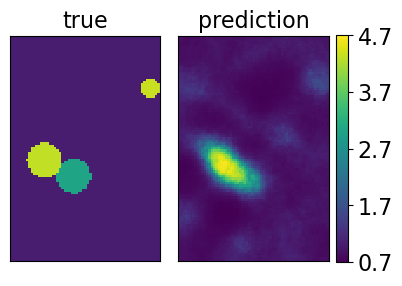}
    \subcaption{\label{fig:app2resulth} (0.359)}
  \end{subfigure}
\caption{\label{fig:app2result} Additional results for CNN2 in increasing order of the scaled error (in brackets) as defined in equation (\ref{eqn:averagescalederror}). True results are on the left and predictions on the right.}  
\end{figure}

\begin{figure}[!h]
  \centering
  \begin{subfigure}[c]{\nhgappwidth}
    \centering    
    \includegraphics[totalheight=\nhgappheight]{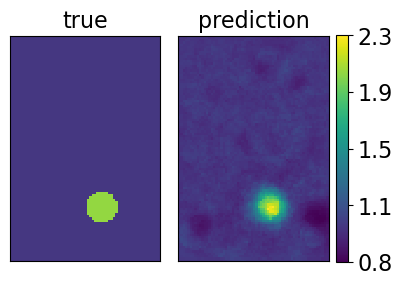}
    \subcaption{\label{fig:app3resulta} (0.0754)}
  \end{subfigure}
  \begin{subfigure}[c]{\nhgappwidth}
    \centering    
    \includegraphics[totalheight=\nhgappheight]{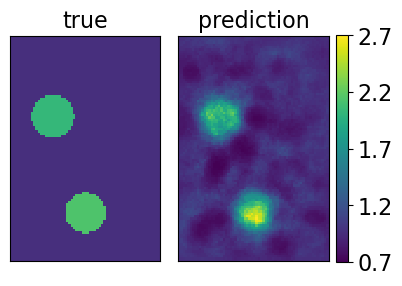}
    \subcaption{\label{fig:app3resultb} (0.0131)}
  \end{subfigure}
  \begin{subfigure}[c]{\nhgappwidth}
    \centering
    \includegraphics[totalheight=\nhgappheight]{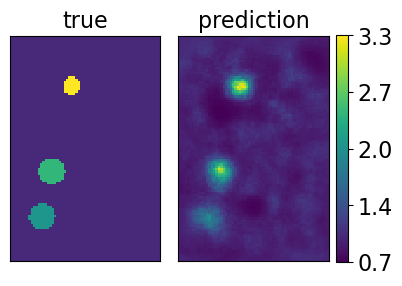}
    \subcaption{\label{fig:app3resultc} (0.161) }
  \end{subfigure}
  \begin{subfigure}[c]{\nhgappwidth}
    \centering
    \includegraphics[totalheight=\nhgappheight]{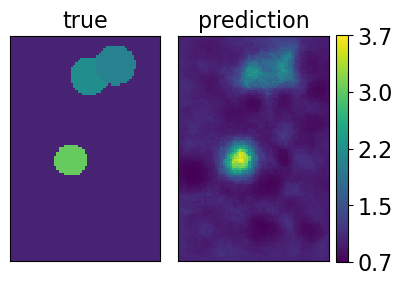}
    \subcaption{\label{fig:app3resultd} (0.181)}
  \end{subfigure}
    \begin{subfigure}[c]{\nhgappwidth}
    \centering    
    \includegraphics[totalheight=\nhgappheight]{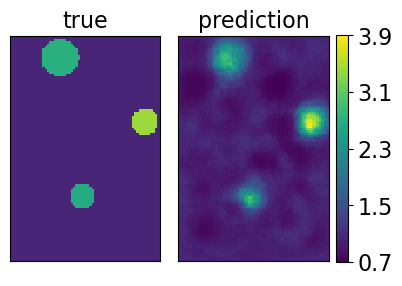}
    \subcaption{\label{fig:app3resulte} (0.199)}
  \end{subfigure}
  \begin{subfigure}[c]{\nhgappwidth}
    \centering    
    \includegraphics[totalheight=\nhgappheight]{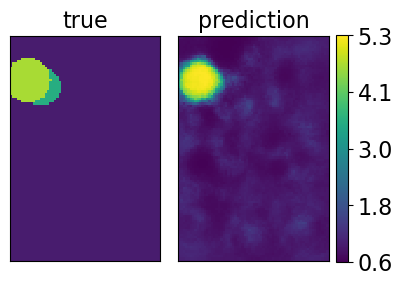}
    \subcaption{\label{fig:app3resultf} (0.223)}
  \end{subfigure}
  \begin{subfigure}[c]{\nhgappwidth}
    \centering
    \includegraphics[totalheight=\nhgappheight]{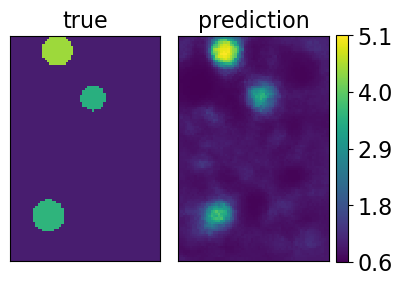}
    \subcaption{\label{fig:app3resultg} (0.249)}
  \end{subfigure}
  \begin{subfigure}[c]{\nhgappwidth}
    \centering
    \includegraphics[totalheight=\nhgappheight]{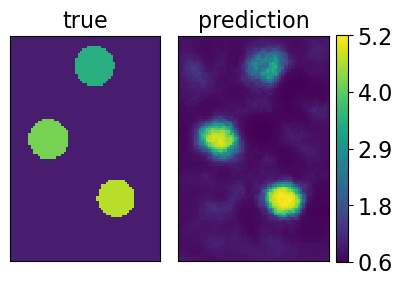}
    \subcaption{\label{fig:app3resulth} (0.305)}
  \end{subfigure}
\caption{\label{fig:app3result} Additional results for CNN3 in increasing order of the scaled error (in brackets) as defined in equation (\ref{eqn:averagescalederror}). True results are on the left and predictions on the right.}  
\end{figure}

We present additional supporting results in figures (\ref{fig:app1result}),(\ref{fig:app2result}) and (\ref{fig:app3result}) in order to provide a better sampling of the results produced by CNN1,CNN2 and CNN3. In general, we see good agreement between true and predicted shear modulus fields. In addition, we see that 1) the location of most inclusions is predicted accurately, 2) the shear modulus of inclusions of small size and low shear modulus is under-predicted, resulting in predictions that can be described best as \textit{wisps} and 3) inclusions close to each other are often combined into one single inclusion in the prediction. 
%
%
\end{appendices}
\clearpage
\bibliography{ei_ml_arxiv2}{}
\bibliographystyle{plain}
\end{document}